\documentclass[twocolumn,prb,superscriptaddress,longbibliography]{revtex4-2}

\usepackage{amsmath,amsfonts,amssymb}
\usepackage{graphicx,color}
\usepackage{hyperref}
\usepackage{float}
\usepackage{multirow}
\usepackage{hhline}
\usepackage{ulem}
\usepackage{mathtools}
\usepackage{siunitx}
\usepackage[caption=false]{subfig}

\newcommand{\ket}[1]{\left|#1\right\rangle}
\newcommand{\bra}[1]{\left\langle#1\right|}
\newcommand{\braket}[3]{\left\langle#1\left|#2\right|#3\right\rangle}

\makeatletter

\@ifundefined{textcolor}{}
{%
 \definecolor{BLACK}{gray}{0}
 \definecolor{WHITE}{gray}{1}
 \definecolor{RED}{rgb}{1,0,0}
 \definecolor{GREEN}{rgb}{0,1,0}
 \definecolor{BLUE}{rgb}{0,0,1}
 \definecolor{CYAN}{cmyk}{1,0,0,0}
 \definecolor{MAGENTA}{cmyk}{0,1,0,0}
 \definecolor{YELLOW}{cmyk}{0,0,1,0}
}

\makeatother

\begin{document}

\title{Autonomous stabilization with programmable stabilized state}

\author{Ziqian Li$^\dag$}
\affiliation{James Franck Institute, University of Chicago, Chicago, Illinois 60637, USA}
\affiliation{Department of Physics, University of Chicago, Chicago, Illinois 60637, USA}
\affiliation{Department of Applied Physics, Stanford University, Stanford, California 94305, USA}
\altaffiliation{These authors contributed equally}

\author{Tanay Roy$^{\dag, *}$}
\affiliation{James Franck Institute, University of Chicago, Chicago, Illinois 60637, USA}
\affiliation{Department of Physics, University of Chicago, Chicago, Illinois 60637, USA}
\altaffiliation{These authors contributed equally}

\author{Yao Lu}
\email{Present address: Superconducting Quantum Materials and Systems Center, Fermi National Accelerator Laboratory (FNAL), Batavia, IL 60510, USA}
\affiliation{James Franck Institute, University of Chicago, Chicago, Illinois 60637, USA}
\affiliation{Department of Physics, University of Chicago, Chicago, Illinois 60637, USA}

\author{Eliot Kapit}
\affiliation{Department of Physics, Colorado School of Mines, Golden, CO 80401}

\author{David I. Schuster}
\affiliation{James Franck Institute, University of Chicago, Chicago, Illinois 60637, USA}
\affiliation{Department of Physics, University of Chicago, Chicago, Illinois 60637, USA}
\affiliation{Pritzker School of Molecular Engineering, University of Chicago, Chicago, Illinois 60637, USA}
\affiliation{Department of Applied Physics, Stanford University, Stanford, California 94305, USA}

\date{\today}
\begin{abstract}

Reservoir engineering is a powerful technique to autonomously stabilize a quantum state. Traditional schemes involving multi-body states typically function for discrete entangled states. In this work, we enhance the stabilization capability to a continuous manifold of states with programmable stabilized state selection using multiple continuous tuning parameters. We experimentally achieve $84.6\%$ and $82.5\%$ stabilization fidelity for the odd and even-parity Bell states as two special points in the manifold. We also perform fast dissipative switching between these opposite parity states within $\SI{1.8}{\micro\second}$ and $\SI{0.9}{\micro\second}$ by sequentially applying different stabilization drives. Our result is a precursor for new reservoir engineering-based error correction schemes.

\end{abstract}
\maketitle


\section{Introduction}

Entanglement is one major resource any quantum protocol utilizes to achieve quantum advantage~\cite{Arute2019, Wu2021}. Generally, the entanglement is created by unitary operations, where dissipation is considered detrimental and should be maximally avoided. Inspired by laser cooling, an alternative approach is to use tailored dissipation for stabilizing entanglement. By coupling the qubit system to some cold reservoirs, one can engineer the Hamiltonian such that the population will flow directionally to the stabilized point in the Hilbert space, and extra entropy is autonomously dumped into the cold reservoir during the process. This provides an extra route to state preparation. In a multiqubit-reservoir coupled system, dissipation engineering can enhance the capabilities of quantum simulation, as predicting the final state of a driven dissipative quantum system is more complex than its unitary counterpart~\cite{PhysRevResearch.2.043042} when all local qubit and reservoir interactions are simultaneously turned on. Dissipation stabilization also inspires autonomous quantum error correction codes (AQEC)~\cite{Verstraete2009, VSLQ2015, Ma2020, Gertler2021, li2023autonomous, li2023hardware} that achieve hardware efficiency in the experiment.

Stabilization has been theoretically proposed and experimentally realized in different platforms, such as superconducting qubits~\cite{Shankar2013,siddiqi2016, Yao2017, 2018Stabilize, Andersen2019, Fu2019, CC2020, Brown2022} and trapped ions~\cite{Lin2013, Leibfried2022}, focusing on stabilizing a single special state per device, such as even or odd parity Bell states. Unlike universal quantum state preparation through unitary gate decomposition, dissipative stabilization requires individual Hamiltonian engineering for each stabilized state through different drive combinations or hardware. This makes the tunable dissipative stabilization a challenging task.  A generalized scheme that allows one to programmatically choose stabilized states from a large class of states per device will expand the toolbox for state preparation. For instance, the ability to choose an arbitrary stabilized state can be used for the implementation of density matrix exponentiation~\cite{Lloyd2014, 2022matrix} by enabling an efficient reset of the input density matrix.


In this work, we realize an autonomous stabilization protocol with superconducting circuits that allows selection from a broad class of states, including the maximally entangled states. We use microwave-only drives with tunable parameters such as drive detunings and strengths that allow fast programmable switching between Bell states of different parities. The system is based on a two-transmon inductive coupler design~\cite{yaothesis2019, Brown2022, roy2022realization, li2023autonomous} that allows fast parametric interactions between qubits without significantly compromising their coherence. The readout resonators are also used as cold reservoirs, eliminating the requirement for extra components. We perform stabilization spectroscopy and demonstrate a fidelity over $78\%$ for all stabilized states. For odd and even parity Bell pairs, we measured $84.6\%$ and $82.5\%$ stabilization fidelity and a stabilization time of $\SI{1.8}{\micro\second}$ and $\SI{0.9}{\micro\second}$ respectively. The structure of the paper is as follows. First, we explain the Hamiltonian construction of the stabilization protocol. Then we discuss the experimental measurement of individual stabilized state and demonstrate a dissipative switch of Bell state parity.

\section{Stabilization theory}

\begin{figure}[t]
    \centering
    \includegraphics[width=\columnwidth]{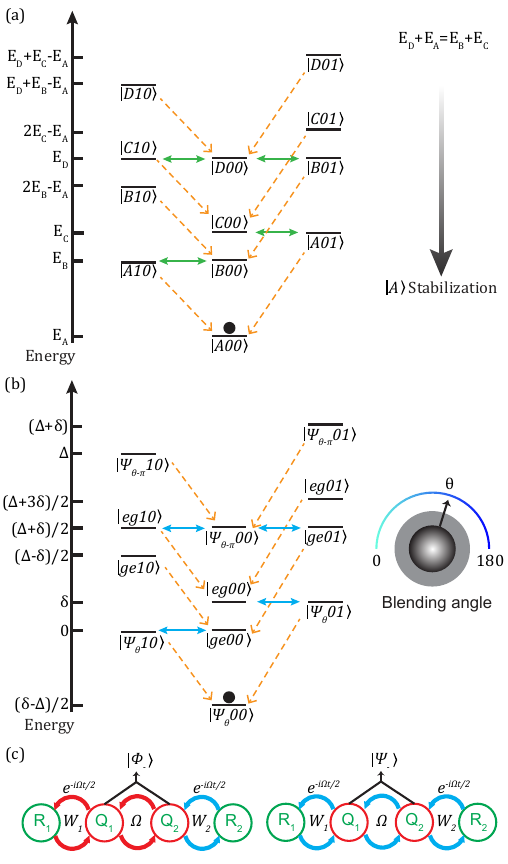}
    \caption{(a) General stabilization scheme. Two qubits' eigenstates $\{\ket{A}, \ket{B}, \ket{C}, \ket{D}\}$ are plotted in the energy level diagram. When the energy relation $E_D+E_A=E_B+E_C$ is satisfied, $\ket{A}$ is stabilized.  Qubit-resonator interactions and resonator photon decay rate $\kappa$ are shown in blue and orange arrows. Qubit decay rate $\gamma$ is assumed slowest and not plotted. (b) Stabilization of entangled states $\ket{\Psi_{\theta}}=\sin\left(\theta/2\right)\ket{gg}-\cos\left(\theta/2\right)\ket{ee}$ or $\ket{\Phi_{\theta}}=\sin\left(\theta/2\right)\ket{ge}-\cos\left(\theta/2\right)\ket{eg}$. (c) A special case of (b) that stabilizes the odd and even parity bell states $\ket{\Phi_{-}}$ and $\ket{\Psi_{-}}$. Circulating arrows are color-coded to represent red (exchange-like) and blue (two-photon-pumping) sidebands respectively. The QQ and QR sideband rates are separate $\Omega$ and $W_j$, and the QR sideband is detuned in frequency by ${\Omega}/{2}$. }
    \centering
    \label{fig:theory}
\end{figure}

We consider a system of two coupled qubit-resonator pairs $\{Q_1, Q_2\}$ and $\{R_1, R_2\}$. The lossy resonators serve as both cold baths and dispersive readouts for the qubits. We label the ground and the first excited states of the qubits $Q_{1/2}$ as $\ket{g}$ and $\ket{e}$, and of the resonators $R_{1/2}$ as $\ket{0}$ and $\ket{1}$, with the full system state being represented as $\ket{Q_1Q_2R_1R_2}$. The system Hamiltonian $H_{\rm sys}=H_{QQ}+H_{QR1}+H_{QR2}$ includes the dominant two-qubit interaction $H_{QQ}$ and qubit-resonator interactions $H_{QRj}, j=\{1,2\}$ acting as perturbations. We label the four eigenstates of $H_{QQ}$ as $\{\ket{A}, \ket{B}, \ket{C}, \ket{D}\}$ with eigenenergies $\{E_A< E_B\leq E_C\leq E_D\}$ so that $\ket{A}$ is the target state to stabilize. Our stabilization scheme involves engineering a one-way flow of population to $\ket{A}$ connecting all intermediate eigenstates of the system.

\begin{figure*}[t]
    \centering
    \includegraphics[width=2\columnwidth]{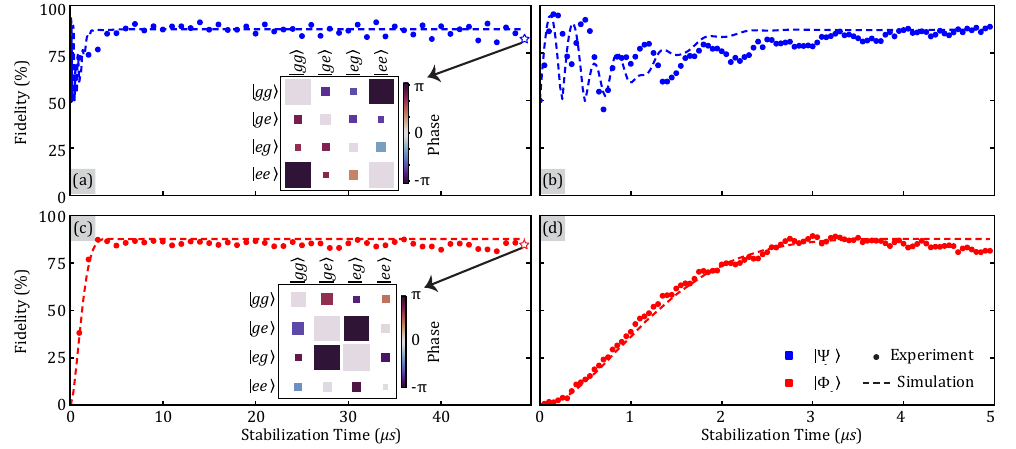}
    \caption{Experimental demonstration of $\ket{\Psi_{-}}$ ((a), (b)) and $\ket{\Phi_{-}}$ ((c), (d)) stabilization with the initial state $\ket{gg}$. Two-qubit state tomography is performed at each time point, and the reconstructed density matrix is used to calculate the target state fidelity. The density matrices reconstructed with 5000 single shot measurements at $\SI{49}{\micro\second}$ are plotted. Lab frame simulation results are shown in dash lines, which matched well in both short and long time scales. Parameters used in simulation: $\{\Omega, W_1, W_2, \Gamma_1, \Gamma_2\}/2\pi=\{2.0, 0.47, 0.47, 0.33, 0.43\}$ MHz for $\ket{\Psi_{-}}$ and $\{3.0, 0.36, 0.36, 0.33, 0.43\}$ MHz for $\ket{\Phi_{-}}$. Qubit coherence time is chosen as $\{T_{1}^{q1}, T_{1}^{q2}, T_{\phi}^{q1}, T_{\phi}^{q2}\}=\{25, 12, 25, 25\} \ \mu$s.}
    \centering
    \label{fig:time_domain}
\end{figure*}

We now derive the energy matching requirements for an efficient stabilization protocol in our two-qubit-two-resonator system depicted in Figure~\ref{fig:theory}(a). We control the form of the target stabilized state $\ket{A}$ by choosing different two-qubit interaction strengths and detunings that control $H_{QQ}$. We change the resonator photon energy in the rotating frame by detuning the QR interactions. The dynamics of $H_{\rm sys}$ are captured by considering the following set of eigenstates: $\{\ket{A},\ket{B},\ket{C},\ket{D}\} \otimes \{\ket{00},\ket{10},\ket{01}\}$. We neglect the resonator state $\ket{11}$ as the probability of simultaneous population in both resonators $\{R_1, R_2\}$ is extremely low when resonator decay rate $\kappa$ is much larger than the qubit decay rate $\gamma$ (assumed identical). The central column in Fig.~\ref{fig:theory}(a) shows the eigenstates of $H_{QQ}$ with no photons in the resonators. The left column represents the same states with one photon in the left ($R_1$) resonator and similarly for the right column is associated with the second resonator ($R_2$). We engineer the photon energies in $R_1$ and $R_2$ to be $E_B-E_A$ and $E_C-E_A$ respectively through tuning the QR interactions $H_{QRj}$. This condition puts two transitions $\ket{A01}\leftrightarrow\ket{C00}$ and $\ket{A10}\leftrightarrow\ket{B00}$ on resonance, shown in Fig.~\ref{fig:theory}(a). If $\bra{A01}H_{QR1}\ket{C00}$ and $\bra{A10}H_{QR2}\ket{B00}$ are non-zero, two on-resonance oscillations between $\ket{C00}$, $\ket{A01}$ and between $\ket{A10}$, $\ket{B00}$ will be created. Since both resonators are lossy, the oscillation will quickly damp to $\ket{A00}$. To complete the downward stabilization path, we need to also connect $\ket{D00}$ into the flow. We further require that the following terms are non-zero so that the transfer path is not blocked: $\bra{B01}H_{QR1}\ket{D00}$, $\bra{C10}H_{QR2}\ket{D00}$. If all four interaction strengths (shown in green double-headed arrows in Fig.~\ref{fig:theory}(a)) are dominant over the qubit decay rate, populations in $\ket{B}$, $\ket{C}$, and $\ket{D}$ will flow to $\ket{A}$. From Fermi's golden rule, the interaction strength between two states is quadratically suppressed by their energy gap and maximized when on-resonance~\cite{Eliot2014}. This imposes a simple energy-matching requirement for efficient stabilization: $E_D+E_A=E_B+E_C$. Energy degeneracy within $\{\ket{B}, \ket{C}, \ket{D}\}$ will not affect the stabilization scheme, because it will not block the dissipative flow to $\ket{A00}$ in Fig.~\ref{fig:theory}(a).

As an explicit demonstration, we first stabilize a continuous set of entangled states $\ket{\Psi_{\theta}}=\sin\left(\theta/2\right)\ket{gg}-\cos\left(\theta/2\right)\ket{ee}$, illustrated in Fig.~\ref{fig:theory}(b). Here, $\theta$ can be regarded as a ``blending angle" between the two even parity states $\ket{gg}$ and $\ket{ee}$. We introduce three sideband transitions into the system: qubit-qubit (QQ) blue sideband $\ket{gg}\leftrightarrow\ket{ee}$ with rate $\Omega$ and two qubit-resonator (QR) blue sidebands $\ket{g0}\leftrightarrow\ket{e1}$ between $Q_{j}$ and $R_{j}$ with rate $W_{j}$. To ensure that $H_{QRj}$ act as perturbations over $H_{QQ}$, we adjust the drive strengths to satisfy $\Omega \gg W_j$. We further detune the QQ, QR1, and QR2 blue sideband by $\delta$, $(\Delta-\delta)/2$, and $(\Delta+\delta)/2$ in frequencies, with $\Delta=\sqrt{\Omega^2+\delta^2}$. The detuning $\delta$ determines the blending angle $\theta=\tan^{-1}\left(\frac{\delta+\Delta}{\Omega}\right)$ with a range of $[0, \frac{\pi}{2})$. In the presence of these three drives, the rotating frame Hamiltonian $H_{\rm sys}$ is
\begin{align}
    {H_{\rm sys}} = & \frac{\Omega}{2}\left(a_{q1}a_{q2}+h.c.\right)+\delta a_{q1}^{\dag}a_{q1} \nonumber\\ 
    & +\frac{W_{1}}{2}\left(a_{q1}a_{r1}+h.c.\right)+\frac{W_{2}}{2}\left(a_{q2}a_{r2}+h.c.\right) \nonumber\\ 
    & +\frac{\Delta+\delta}{2}a_{r1}^{\dag}a_{r1}+\frac{\Delta-\delta}{2}a_{r2}^{\dag}a_{r2}.
    \label{eq:detuning_even}
\end{align}
Here $a_{qj}$ and $a_{rj}$ are separately the $j$-th qubit's and resonator's annihilation operator. Under the combined conditions $\Omega\gg W_{j}\sim\kappa\gg\gamma$ and $W_j=W$, the eigenstates with zero resonator photons are $\{\ket{\Psi_{\theta}00}, \ket{ge00}, \ket{eg00}, \ket{\Psi_{\pi-\theta}00}\}$, with corresponding eigenenergies $\{\left(\delta-\Delta\right)/2,0,\delta,\left(\delta+\Delta\right)/2\}$. Assuming the lossy resonator has a Lorentzian energy spectrum, the two-step refilling rate $\Gamma_t$ from $\ket{eg00}$ to $\ket{\Psi_{\theta}00}$ ($\ket{eg00}\leftrightarrow\ket{\Psi_{\theta}01}$, $\ket{\Psi_{\theta}01}\rightarrow\ket{\Psi_{\theta}00}$) is~\cite{Eliot2014} 
\begin{align}
\Gamma_t=\frac{W^2\cos^2\left(\theta/2\right)\kappa}{\kappa^2+W^2\cos^2\left(\theta/2\right)}.
\label{eq:refilling}
\end{align}
The other two-step transitions $\ket{ge00}\rightarrow\ket{\Psi_{\theta}00}$, $\ket{\Psi_{\theta-\pi}00}\rightarrow\ket{ge00}$, and $\ket{\Psi_{\theta-\pi}00}\rightarrow\ket{eg00}$ also have the same rate. Therefore, the steady-state fidelity $\mathcal{F}_{\infty}$ for $\ket{\Psi_{\theta}00}$ is (ignoring all off-resonant transitions, see Appendix.~\ref{app:ss_fidelity} for detail)
\begin{align}
\mathcal{F}_{\infty}=\left(\frac{\Gamma_t+\gamma\sin^2\left(\theta/2\right)}{\Gamma_t+\gamma}\right)^2.
\label{eq:fidelity}
\end{align}

Similarly, we can stabilize another set of entangled states with odd parity $\ket{\Phi_{\theta}}=\sin\left(\theta/2\right)\ket{ge}-\cos\left(\theta/2\right)\ket{eg}$. We introduce three sideband interactions: QQ red $\ket{eg}\leftrightarrow\ket{ge}$, QR1 red  $\ket{e0}\leftrightarrow\ket{g1}$, and QR2 blue $\ket{g0}\leftrightarrow\ket{e1}$ with rates $\{\Omega, W_3, W_4\}$ and frequency detunings $\{\delta, (\Delta+\delta)/2, (\Delta-\delta)/2\}$ respectively. Under this condition, four resonant interactions will appear: $\ket{gg00}\leftrightarrow\ket{\Phi_{\theta}01}$, $\ket{ee00}\leftrightarrow\ket{\Phi_{\theta}10}$, $\ket{ee01}\leftrightarrow\ket{\Phi_{\theta-\pi}00}$, and $\ket{gg10}\leftrightarrow\ket{\Phi_{\theta-\pi}00}$. The detuning similarly sets the blending angle $\theta=\arctan\left(\frac{\delta+\Delta}{\Omega}\right)$.

With the above construction, we create a stabilization protocol that can freely tune the blending angles. As a special case, when QQ sideband detuning $(\delta=0)$, the blending angle for both cases is $\theta=\frac{\pi}{2}$, which corresponds to the odd and even parity Bell states $\ket{\Phi_{-}}=(\ket{ge}-\ket{eg})/\sqrt{2}$ and $\ket{\Psi_{-}}=(\ket{gg}-\ket{ee})/\sqrt{2}$, shown in Fig.~\ref{fig:theory}(c). 

In fact, this stabilization protocol can be generalized to stabilize an even larger group of states, including both entangled and product states, as long as the energy matching requirement $E_D+E_A=E_B+E_C$ is satisfied when engineering $H_{QQ}$. The following is a list of tunable parameters to engineer $H_{QQ}$: QQ sideband strength $\Omega$, QQ sideband detunings $\delta$, single qubit Rabi drive strength, and single qubit Rabi drive detunings. Corresponding stabilized state $\ket{A}$ is determined from $H_{QQ}$. Details about the stabilizable manifold are discussed in Appendix.~\ref{app:other}.



\section{Experimental results}

We perform the stabilization experiment in a system with two transmons capacitively coupled to two lossy resonators (See Appendix.~\ref{app:device}). Two transmons are inductively coupled through a SQUID loop. All QQ sidebands and QR red sidebands are realized through modulating the SQUID flux at corresponding transition frequencies. QR blue sidebands are achieved by sending a charge drive to the transmon at half the transition frequencies. The experimentally measured qubit coherence are $T_1=\SI{24.3}{\micro\second}\ (\SI{9.1}{\micro\second})$, $T_{\rm Ram}=\SI{15.2}{\micro\second}\ (\SI{9.8}{\micro\second})$, $T_{\rm echo}=\SI{24.6}{\micro\second}\ (\SI{14.3}{\micro\second})$ for $Q_{1}(Q_{2})$, and the measured resonator decay rate $\kappa/2\pi$ are $\{0.33, 0.43\}$ MHz for $R_1$ and $R_2$ respectively.

\begin{figure}[t]
    \centering
    \includegraphics[width=\columnwidth]{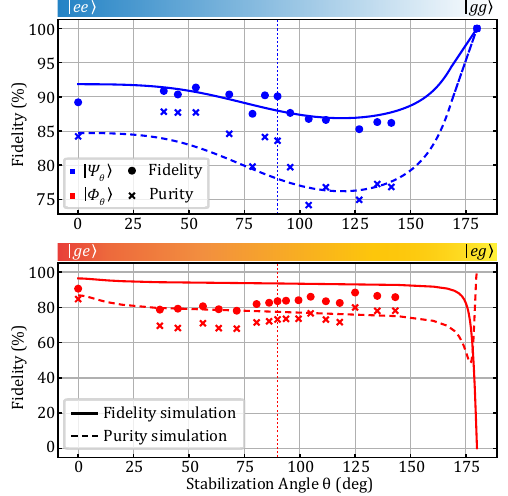}
    \caption{Spectroscopy of universal Bell-state stabilization. $\ket{\Psi_{\theta}}$ (top) and $\ket{\Phi_{\theta}}$ (bottom) are separately stabilized with a measured fidelity above $78\%$ among different blending angle $\theta$. The fidelities are measured after $\SI{40}{\micro\second}$ of stabilization. For stabilizing $\ket{gg}$, no external drives are applied. For $\ket{\Phi_{\theta}}$ case, the fidelity dropped to $0$ near $\theta=\pi$. The dotted lines indicate simulated fidelities for the odd and even parity Bell state stabilization. All parameters used in the simulation are the same as in Fig.~\ref{fig:time_domain}. }
    \centering
    \label{fig:universal}
\end{figure}

Figure~\ref{fig:time_domain} shows the time evolution of state fidelity for the odd and even parity Bell state stabilization. To stabilize $\ket{\Psi_{-}}$, a $\Omega=2\pi\times2.0$ MHz QQ blue sideband, $W_1=W_2=2\pi\times0.47$ MHz QR blue sidebands are simultaneously applied to the system. Both QR sidebands are detuned by $\Omega/2=2\pi\times1.0$ MHz in frequency to implement the stabilization scheme depicted in Fig.~\ref{fig:theory}(c). For each stabilization experiment, we reconstruct the system density matrix through two-qubit state tomography using 5000 repetitions of 9 different pre-rotations. The stabilization fidelity measured at $\SI{49}{\micro\second}$ (much longer than single qubit $T_1$ and $T_{\rm Ram}$) is $82.5\%$. To stabilize $\ket{\Phi_{-}}$, a $\Omega=2\pi\times3.0$ MHz QQ red sideband, $W_1=W_2=2\pi\times0.36$ MHz QR1 red and QR2 blue sidebands are simultaneously applied to the system, with both QR sidebands detuned by $\Omega/2=2\pi\times1.5$ MHz. The stabilization fidelity measured at $\SI{49}{\micro\second}$ is $84.6\%$. The two-qubit state tomography data at $\SI{49}{\micro\second}$ after ZZ coupling correction~\cite{ZZcorrection} are shown for both stabilization cases. 

Next, we introduce QQ sideband detunings $\delta$ and stabilize more general entangled states $\ket{\Psi_{\theta}}$ and $\ket{\Phi_{\theta}}$. We choose the same sideband strengths ($\{\Omega, W_1, W_2\}/2\pi=\{2.0, 0.47, 0.47\}(\{3.0, 0.36, 0.36\})$ MHz for $\ket{\Psi_{\theta}}$($\ket{\Phi_{\theta}}$) case) and detune QR sideband frequencies accordingly to maximize the stabilization fidelity measured at $\SI{40}{\micro\second}$. The experimentally measured state fidelity and state purity as a function of $\theta$ are shown in Fig.~\ref{fig:universal}. Under the current QR sideband color combination, $\ket{\Phi_{\theta}}$ fails to stabilize near $\theta=180^\circ$. This is because the interaction strength $\bra{gg00}H_{\rm sys}\ket{\Phi_{\theta}01}$ and $\bra{ee00}H_{\rm sys}\ket{\Phi_{\theta}10}$ are close to $0$. Swapping QR1 and QR2 sidebands' color and detuning performs a transformation $\theta\rightarrow\theta-\pi$ in the stabilized state. This ensures a high stabilization fidelity for arbitrary stabilization angles. Details about changing sideband colors and detunings to ensure high fidelity are presented in Appendix.~\ref{app:color}

\begin{figure}[t]
    \centering
    \includegraphics[width=\columnwidth]{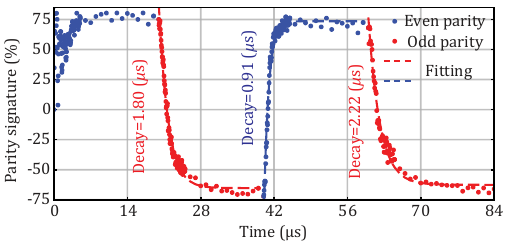}
    \caption{Dissipative switching of Bell state parity. The initial state is $\ket{gg}$, and the switch status is set to even parity between $\left[\SI{0}{\micro\second}, \SI{20}{\micro\second}\right]$ and $\left[\SI{40}{\micro\second}, \SI{60}{\micro\second}\right]$, and to odd parity between $\left[\SI{20}{\micro\second}, \SI{40}{\micro\second}\right]$ and $\left[\SI{60}{\micro\second}, \SI{85}{\micro\second}\right]$. Each experimental point is measured with the two-qubit state tomography. Stabilization time is calculated by fitting the parity signature to exponential decay after each switching event.}
    \centering
    \label{fig:switch_plot}
\end{figure}

The flexibility in our schemes and easy access to different sidebands in our device allow a further demonstration --- fast dissipative switching between stabilized states. Here, we implement such an operation that can flip the parity of the stabilized Bell pair by changing sideband combinations, shown in Fig.~\ref{fig:theory}. To quantify the stabilized parity, we measure the system's density matrix $\rho$ and define the parity signature as $2(|\braket{ee}{\rho}{gg}|-|\braket{ge}{\rho}{eg}|)$ describing the difference in relevant coherence parameters. The results are shown in Fig.~\ref{fig:switch_plot}. The scaling factor is chosen such that the ideal even and odd Bell pairs have parity signatures of $\pm1$. Starting from the ground state $\ket{Q_1Q_2}=\ket{gg}$, the stabilized state is set to even parity Bell pair $(\ket{gg}-\ket{ee})/\sqrt{2}$, and we switch the parity every $\SI{20}{\micro\second}$. At $\SI{20}{\micro\second}$, the stabilized state is switched to odd parity Bell pair $(\ket{ge}-\ket{eg})/\sqrt{2}$, and stabilization happens quickly with a time constant $\tau_{r}=\SI{1.80}{\micro\second}$. At $\SI{40}{\micro\second}$, the switching from odd to even parity results in a faster stabilization with $\tau_{b}=\SI{0.91}{\micro\second}$. The switching at $\SI{60}{\micro\second}$ to odd Bell state shows a similar $\tau_{r}$ of $\SI{2.20}{\micro\second}$. We leave the stabilization drives turned on for another $\SI{25}{\micro\second}$ to prove that the performance is not degraded after a few switching operations.

Further improvement of the stabilized state's fidelity is possible by reducing the transition ratio $\frac{\gamma}{\Gamma_t}$ (from Eq.~\ref{eq:fidelity}) and increasing QQ sideband rate $\Omega$ for a larger energy gap. Increasing qubit dephasing time also improves stabilization fidelity (discussed in the Appendix.~\ref{app:robust}). To speed up the stabilization, i.e., reduce time constants, we need to increase the refilling rate $\Gamma_t$. Since QR sideband rate $W$ is bounded by the QQ sideband rate $\Omega$ to ensure the validity of the perturbative approximation, given a fixed $W$, $\Gamma_t$ is maximized when the resonator decay rate $\kappa=W\cos(\theta/2)$. For the even and odd parity Bell states, further increase in both resonators' $\kappa$ compared to our current parameters would thus be beneficial. More details about stabilization robustness are discussed in Appendix.~\ref{app:robust}. To stabilize a more general set of states shown in Appendix.~\ref{app:other}, longer qubit coherence is needed to improve the experimental resolution between different stabilized states in this manifold and is a subject of future work.


\section{Conclusion}
In conclusion, we demonstrate a two-qubit programmable stabilization scheme that can autonomously stabilize a continuous set of entangled states. We develop an inductively coupled two-qubit device that provides access to both QQ and QR sideband interactions required. The stabilization fidelity among all stabilization angles is above $78\%$, specifically, we achieved high Bell pair stabilization fidelity ($84.6\%$ for the odd parity and $82.5\%$ for the even parity) as two special points. We further demonstrate a parity switching capability between the Bell pairs with fast stabilization time constants ($< 2 \ \mu$s). We believe such freedom in choosing stabilized states will inspire generalization to autonomous stabilization of larger systems, large-scale many-body entanglement~\cite{PhysRevResearch.2.043042}, remote entanglement~\cite{Ma_2021}, density matrix exponentiation~\cite{Lloyd2014, 2022matrix}, and new AQEC logical codewords in the future.


\section{Acknowledgment}
This work was supported by AFOSR Grant No.
FA9550-19-1-0399 and ARO Grant No. W911NF-17-S0001. Devices are fabricated in the Pritzker Nanofabrication Facility at the University of Chicago, which receives
support from Soft and Hybrid Nanotechnology Experimental (SHyNE) Resource (NSF ECCS-1542205), a node
of the National Science Foundation’s National Nanotechnology Coordinated Infrastructure. This work also made use of the shared facilities at the University of Chicago Materials Research Science and Engineering Center, supported by the National Science Foundation under award number DMR-2011854. EK's research was additionally supported by NSF Grant No. PHY-1653820.

\clearpage
\newpage

\appendix

\section{Device Hamiltonian and realized sidebands}
\label{app:device}
\begin{figure}[b]
    \centering
    \includegraphics[width=\columnwidth]{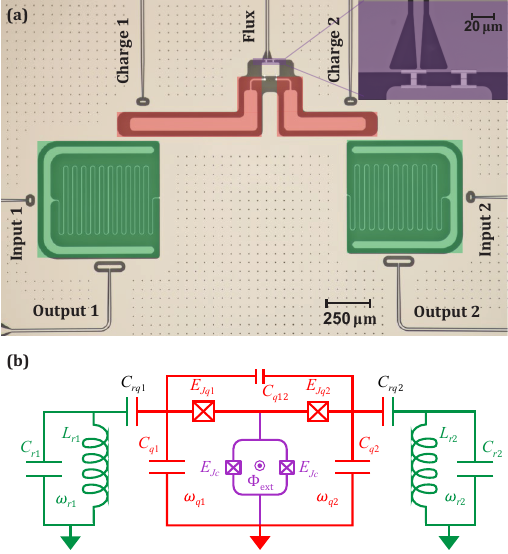}
    \caption{The device. (a) False-colored optical image of the device. The inset shows a zoomed-in image of the inductive coupler. (b) Circuit diagram of the device. Coupler junctions' intrinsic capacitance $C_{qc}$ is included in the quantization analysis.}
    \centering
    \label{suppfig:device}
\end{figure}

Figure~\ref{suppfig:device} shows the inductively coupled two-qubit device used in the stabilization demonstration. Two transmon qubits (red) share a common ground, which is interrupted by a SQUID (purple). Each transmon is capacitively coupled to a lossy resonator (green) also serving as the dispersive readout.

\begin{table}[h]
        \begin{tabular}{cccc}
		    \hline\hline
		     $\varphi_{\rm dc}=0.0$ & $T_{1}$$\left(\mu s\right)$ & $T_{\rm Ram}$$\left(\mu s\right)$ & $T_{\rm echo}$$\left(\mu s\right)$ \\ \hline\hline
		    $Q_{1}$ & $31.6$ & $28.4$ & $26.6$ \\  \hline
		    $Q_{2}$ & $2.8$ & $4.9$ &  \\  \hline\hline
		\end{tabular}
		
		\begin{tabular}{cccc}
		    \hline
		    \hline
		     $\varphi_{\rm dc}=0.3795\pi$ & $T_{1}$$\left(\mu s\right)$ & $T_{\rm Ram}$$\left(\mu s\right)$ & $T_{\rm echo}$$\left(\mu s\right)$ \\  \hline\hline
		    $Q_{1}$ & $24.3$ & $15.2$ & $24.6$ \\  \hline
		    $Q_{2}$ & $9.1$ & $9.8$ & $14.3$ \\  \hline
		    $R_{1}$ & $0.48$ & & \\ \hline
		    $R_{2}$ & $0.37$ & & \\ \hline\hline
		\end{tabular}
		\label{table:qutrit_coherence}
		
		\begin{tabular}{ccc}
		    \hline
		    \hline
		     Parameter & Symbol & Value/$2\pi$ \\  \hline\hline
		     $Q_{1}$ ge frequency & $\omega_{q1}/2\pi$ & $3.2046$ (GHz)  \\  \hline
		     $Q_{2}$ ge frequency & $\omega_{q2}/2\pi$ & $3.6624$ (GHz)  \\  \hline
		     $Q_{1}$ anharmonicity & $\alpha_{1}/2\pi$ & $-116.3$ (MHz)  \\  \hline
		     $Q_{2}$ anharmonicity & $\alpha_{2}/2\pi$ & $-159.5$ (MHz)  \\  \hline
		     Readout1 frequency & $\omega_{r1}/2\pi$ & $4.9946$ (GHz)  \\  \hline
		     Readout2 frequency & $\omega_{r2}/2\pi$ & $5.4505$ (GHz)  \\  \hline
		     $\left(E_{\ket{ee}}-E_{\ket{ge}}\right)-\left(E_{\ket{eg}}-E_{\ket{gg}}\right)$ & $ZZ/2\pi$ & -261 (kHz) \\ \hline
             Readout1 fidelity & & $88.87\%$ \\ \hline
             Readout2 fidelity & & $81.76\%$ \\ \hline\hline
		\end{tabular}
		
		\caption{Device coherence and frequencies at the coupler biasing point $\varphi_{\rm dc}=0.3795\pi$ (top and middle), and the coherence at the coupler sweet spot $\Phi_{\rm dc}=0\pi$ (bottom). $Q_2$ experiences higher loss at the sweet spot from a near two-level system. At the sweet spot, $Q_1$'s decoherence and echo time are similar to those at our experiment's biasing point.}
		\label{table:frequency}
\end{table}

The flux line near the SQUID provides a continuous DC bias $\varphi_{\rm dc}=0.3795\pi$ in our experiment. The measured qubit coherence and frequencies at this flux point are shown in Table.~\ref{table:frequency}. By sending RF flux drives at appropriate frequencies through the flux line, the inductive coupler can provide either QQ red sideband or QQ blue sideband. The Hamiltonian for qubits and coupler is shown in Eq.~\ref{eq:H1}. For a detailed description on the adiabatic approximation of the inductive coupler, we draw the attention of the readers to the references \onlinecite{Yao2017} and \onlinecite{li2023autonomous}. Following is a summary. The Hamiltonian of the circuit containing two qubits and the coupler is described by
\begin{subequations}
\begin{align}
    H = & \overrightarrow{n}^{\intercal}C_{L}^{-1}\overrightarrow{n} -E_{j1}\cos{\left(\phi_{c}-\phi_{1}\right)}-E_{j2}\cos{\left(\phi_{2}-\phi_{c}\right)}, \nonumber\\
    & -E_{jc}\cos{\left(\frac{\Phi_{\rm ext}}{\Phi_{0}}\right)}\cos{\left(\phi_{c}\right)}, \label{eq:H1} \\
    C_{L} = &
    \begin{bmatrix}
               C_{q1}+C_{q12} & -C_{q12} & 0 \\
               -C_{q12} & C_{q2}+C_{q12} & 0 \\
               0 & 0 & C_{q1}+C_{q2}+C_{qc}
    \end{bmatrix} \label{eq:H3} \\
    \overrightarrow{n}^{\intercal} = & \left(n_{1}, n_{2}, n_{c}\right). \nonumber
\end{align}
\end{subequations}

Here, phase variables $\phi_{1,2,c}$, charge variables $n_{1,2,c}$, and Josephson energy $E_{j1, j2, jc}$ are for qubit 1, qubit 2, and the coupler, and $C_{qc}$ is the intrinsic capacitance from the coupler junction. After quantization, one can arrive at the following:
\begin{subequations}
\begin{align}
     H_{ab} = & \omega_{q1}a_{q1}^{\dagger}a_{q1}+\omega_{q2}a_{q2}^{\dagger}a_{q2} \nonumber\\
     & +\frac{\alpha_{1}}{2}a_{q1}^{\dagger}a_{q1}^{\dagger}a_{q1}a_{q1}+\frac{\alpha_{2}}{2}a_{q2}^{\dagger}a_{q2}^{\dagger}a_{q2}a_{q2} \nonumber\\
     & +g_{1}\left(t\right)\left(a_{q1}^{\dagger}+a_{q1}\right)\left(a_{q2}^{\dagger}+a_{q2}\right) \nonumber\\
     & +g_{2}\left(-a_{q1}^{\dagger}+a_{q1}\right)\left(-a_{q2}^{\dagger}+a_{q2}\right), \label{eq:adb}\\
     g_{1}\left(t\right) = & \frac{\sqrt{E_{j1}E_{j2}}} {2E_{jc}\cos{\left(\frac{\Phi_{\rm ext}\left(t\right)}{\Phi_0}\right)}} \sqrt{\omega_{q1}\omega_{q2}} \label{eq:quantization}, \\
     g_{2} = & \frac{\sqrt{C_{q1}C_{q2}}}{2C_{q12}} \sqrt{\omega_{q1}\omega_{q2}}. 
\end{align}
\end{subequations}

\begin{figure}[t]
    \centering
    \includegraphics[width=\columnwidth]{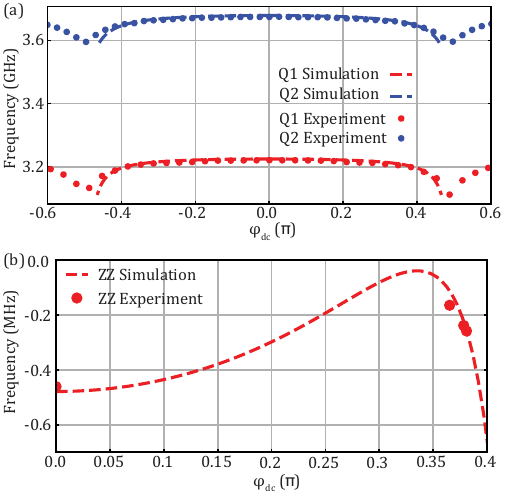}
    \caption{Coupler DC flux sweep of (a) qubits frequencies and (b) ZZ coupling strength between qubits. Circuit quantization results and experimentally measured data are separately shown in dash lines and dots.}
    \centering
    \label{suppfig:twotone}
\end{figure}
In the experiment, we have $E_{jc}=\SI{1700}{\giga\hertz}\gg E_{j1}(E_{j2})=\SI{12.2}{\giga\hertz}(\SI{12.3}{\giga\hertz})$, making the coupler mode much heavier than the qubit modes. This strong asymmetry enables the adiabatic removal of the coupler dynamics: We treat the coupler as a linear inductance and assume the coupler mode is static. By removing the linear mode through minimizing the system energy, we arrive at the approximated Hamiltonian in Eq.~\ref{eq:adb}. When modulating the external RF flux threaded the coupler, the transverse coupling strength $g_1(t)$ will also be modulated accordingly. By plugging the RF flux modulation $\Phi_{\rm ext}/\Phi_{0}=\varphi_{\rm dc}+\epsilon\cos\left(\omega_{d}t\right)$ into Eq.~\ref{eq:quantization} and assuming the flux modulation $\epsilon$ is much smaller than $\Phi_{\rm dc}$, we have:
\begin{subequations}
\begin{align}
     g_{1}\left(t\right) = & \frac{\sqrt{E_{j1}E_{j2}}}{2E_{jc}}\sqrt{\omega_{q1}\omega_{q2}}\frac{1}{\cos\left(\varphi_{\rm dc}+\epsilon\cos\left(\omega_{d}t\right)\right)} \nonumber \\
     \approx & \frac{\sqrt{E_{j1}E_{j2}}}{2E_{jc}}\sqrt{\omega_{q1}\omega_{q2}}\frac{\left(1+\epsilon\sin\left(\omega_{d}t\right)\tan\left(\varphi_{\rm dc}\right)\right)}{\cos\left(\varphi_{\rm dc}\right)}
     \label{eq:anl rate}
\end{align}
\end{subequations}
Therefore the QQ sideband rate is roughly $\epsilon\frac{\sqrt{E_{j1}E_{j2}}}{2E_{jc}}\sqrt{\omega_{q1}\omega_{q2}}\frac{\tan\left(\varphi_{\rm dc}\right)}{\cos\left(\varphi_{\rm dc}\right)}$, which is proportional to the flux modulation rate and increases with $\varphi_{\rm dc}$. We choose $\varphi_{\rm dc}=0.3795\pi$ based on qubit coherence, sideband rate, and small ZZ coupling between qubits. When the resonator is capacitively coupled to the qubit at strength $g_{qr}$ and frequency difference $\Delta$, the QR sideband can also be modulated through the flux line, and the sideband rate is multiplied by an extra mode dressing coefficient $\frac{g_{qr}}{\Delta}$. By changing the RF modulation frequency $\omega_{d}$, we can activate different colors of QQ and QR sidebands.

Both QQ red, QQ blue, and QR red sidebands are generated through flux modulation at a decent rate. The QR blue sidebands have a different choice to activate~\cite{wallraff2007sidband}: The direct charge drives at half of the transition frequency with amplitude $\epsilon_{q}$ can provide an effective rate of $W={16g_{qr}^{3}\epsilon_{q}^{2}}/{\Delta^4}$. This turned out to be easier to realize with our device, and we achieve at least $\SI{0.5}{\mega\hertz}$ QR blue sideband rate.

\begin{figure}[t]
    \centering
    \includegraphics[width=\columnwidth]{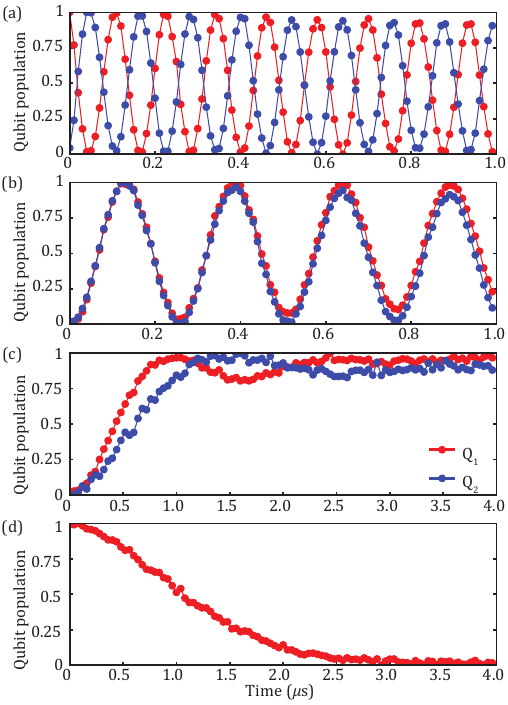}
    \caption{Experimentally realized QQ and QR sidebands. From top to bottom are separately (a) QQ red sideband $\ket{ge}\leftrightarrow\ket{eg}$, (b) QQ blue sideband $\ket{gg}\leftrightarrow\ket{ee}$, (c) QR blue sideband between QR1 and QR2 $\ket{g0}\leftrightarrow\ket{e1}$, and (d) QR1 red sideband $\ket{e0}\leftrightarrow\ket{g1}$. Readout on $Q_1$ (red) and $Q_2$ (blue) are scaled between $0$ ($\ket{g}$) and $1$ ($\ket{e}$). Data points are connected for visual guidance.}
    \centering
    \label{suppfig:rate}
\end{figure}
Figure~\ref{suppfig:rate} demonstrates all realized QQ and QR sidebands needed for the stabilization experiments. Fast QQ red sidebands at $\SI{8.5}{\mega\hertz}$ and modest QQ blue sidebands at $\SI{3.9}{\mega\hertz}$ are performed in the experiment through RF flux modulation of the inductive coupler. Both readouts for QQ red sideband (with initial state $\ket{eg}$) and QQ blue sideband (with initial state $\ket{gg}$) are shown in Fig.~\ref{suppfig:rate}(a) and (b). The QR blue sidebands are generated through the charge lines that are coupled to the qubit pads, shown in Fig.~\ref{suppfig:rate}(c) with initial state $\ket{g0}$. The QR1 red sidebands are activated through the coupler flux modulation. The on-resonance readout trace for $Q_{1}$ starting at $\ket{e0}$ is plotted in Fig.~\ref{suppfig:rate}(d). 

\section{Derivation of steady-state fidelity}
\label{app:ss_fidelity}
 We take the stabilization of $\ket{\Psi_{\theta}}=\sin\left(\theta/2\right)\ket{gg}-\cos\left(\theta/2\right)\ket{ee}$ as an example to compute the steady-state fidelity in detail. Suppose the steady state population at the four basis states $\{\ket{\Psi_{\theta}}, \ket{ge}, \ket{eg}, \ket{\Psi_{\theta-\pi}}\}$ are separately $\{w,x,y,z\}$. We assume the photon population in both resonators are transitional and ignore their contribution to the steady-state fidelity. This means $w+x+y+z=1$. The steady-state configuration should balance the following two processes: 
 
(a) two-step refilling process: $\ket{ge}\rightarrow\ket{\Psi_{\theta}}$, $\ket{eg}\rightarrow\ket{\Psi_{\theta}}$, $\ket{ge}\rightarrow\ket{\Psi_{\theta}}$, and $\ket{ge}\rightarrow\ket{\Psi_{\theta}}$. All the transition rates are the same 
 \begin{align}
 \Gamma_{t}&=\frac{W^2\cos^2\left(\theta/2\right)\kappa}{\kappa^2+W^2\cos^2\left(\theta/2\right)}
 \end{align}
 (b) Single photon loss in each qubit. The following four transitions have the same rate $\sin^2\left(\theta/2\right)\gamma$: $\ket{ge}\rightarrow\ket{\Psi_{\theta}}$, $\ket{eg}\rightarrow\ket{\Psi_{\theta}}$, $\ket{ge}\rightarrow\ket{\Psi_{\theta-\pi}}$, and $\ket{eg}\rightarrow\ket{\Psi_{\theta-\pi}}$. The reversed four transitions have the same rate $\cos^2\left(\theta/2\right)\gamma$.

Therefore, the steady-state population should satisfy the following equations
\begin{equation}
    \begin{cases}
      \left(\Gamma_t+\sin^2\left(\theta/2\right)\gamma\right)(x+y)-2\cos^2\left(\theta/2\right)\gamma w &= 0,\\
      \left(\Gamma_t+\sin^2\left(\theta/2\right)\gamma\right)z+\cos^2\left(\theta/2\right)\gamma w-(\Gamma_t+\gamma)x &= 0,\\
      \left(\Gamma_t+\sin^2\left(\theta/2\right)\gamma\right)z+\cos^2\left(\theta/2\right)\gamma w-(\Gamma_t+\gamma)y &= 0,\\
      2\cos^2\left(\theta/2\right)\gamma (x+y)-\left(2\Gamma_t+2\sin^2\left(\theta/2\right)\gamma\right)z &= 0,\\
      w+x+y+z &= 1. \\
    \end{cases}       
\end{equation}
This gives the following populations
\begin{equation}
\begin{cases}
w &= \left(\frac{\Gamma_t+\gamma\sin^2\left(\theta/2\right)}{\Gamma_t+\gamma}\right)^2 ,\\
x &= \frac{\cos^2\left(\theta/2\right)\gamma}{\Gamma_t+\gamma-\cos^2\left(\theta/2\right)\gamma}w ,\\
y &= x ,\\
z &= \frac{\cos^2\left(\theta/2\right)\gamma}{\Gamma_t+\sin^2\left(\theta/2\right)\gamma}x .\\
\end{cases}
\end{equation}
And $w=\mathcal{F}_{\infty}$ is the steady state fidelity. 

\section{Stabilization robustness}
\label{app:robust}
\begin{figure}[t]
    \centering
    \includegraphics[width=\columnwidth]{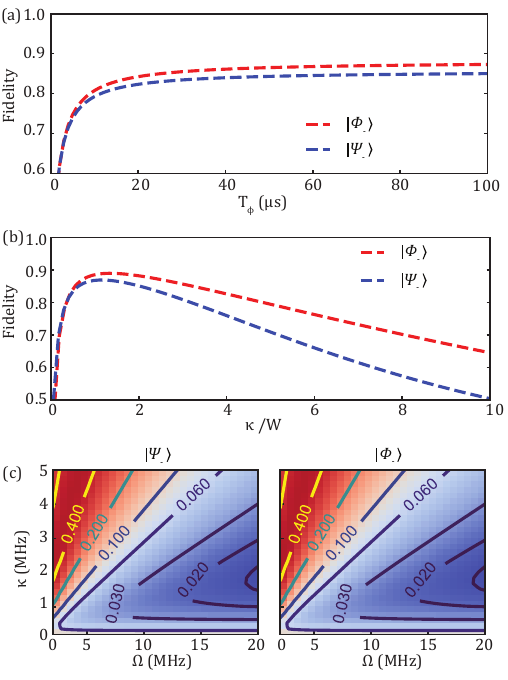}
    \caption{Rotating frame simulation of even and odd parity bell states' stabilization fidelity: (a) Sweeping qubit dephasing time. The other simulation parameters are the same as the experiments. (b) Sweeping the ratio between resonator decay rate $\kappa$ and QR sideband strength $W$ without qubit dephasing. (c) 2D sweep of $\kappa$ and QQ sideband strength $\Omega$ without qubit dephasing, setting $W=\kappa$. Infidelities are shown on the contours.}
    \centering
    \label{suppfig:stable}
\end{figure}
We study the stabilization robustness for $\ket{\Psi_{-}}$ and $\ket{\Phi_{-}}$ in this section. For other stabilization angles, the discussion is similar. Figure~\ref{suppfig:stable} shows the rotating frame simulation of steady-state fidelity by sweeping different stabilization parameters. For the $\ket{\Psi_{-}}$ case, the Hamiltonian used in the simulation is Eq.~\ref{eq:detuning_even}, and for the $\ket{\Phi_{-}}$ case the Hamiltonian is modified accordingly with a different sideband combinations (See Fig.~\ref{fig:theory}(c)). We study the state fidelity by varying parameters step-by-step towards the ideal case. First, we show that longer qubit dephasing time helps improve steady-state fidelity. In Fig.~\ref{suppfig:stable} (a), we sweep qubit's dephasing time (assuming the same for both qubits) while choosing the following parameters $\{\Omega, \frac{W_1}{2}, \frac{W_2}{2}, \Gamma_1, \Gamma_2\}/2\pi$ in the simulation
\begin{subequations}
\begin{align}
\ket{\Psi_{-}} \text{case}:& \{1.4, 0.35, 0.35, 0.30, 0.33\} \text{MHz}, \nonumber\\
\ket{\Phi_{-}} \text{case}:& \{3.0, 0.32, 0.32, 0.30, 0.33\} \text{MHz}. \nonumber
\end{align}
\end{subequations}
We set the following qubit decoherence time $\{T_1^{q1}, T_1^{q2}\}=\{21, 9\}\mu$s. The steady state fidelities rise above $80\%$ quickly after $T_{\phi}$ exceeds $\SI{10}{\micro\second}$. The fidelity for odd and even parity bell pairs saturate at $87.3\%$ and $85.0\%$ with the parameters used in the simulation. This demonstrates that steady-state fidelity increases as qubit dephasing time increases.

In Fig.~\ref{suppfig:stable}(b), we ignore the qubit dephasing and only sweep resonator decay rate $\kappa$. For simplicity, we assume QR sideband rates and resonator decay rate are the same: $W_1=W_2=W$ and $\kappa_{1}=\kappa_{2}=\kappa$. The fidelity peak for both parity pairs appears at $W=\kappa$. This can be understood as the refilling rate $\Gamma_{t}$ (Eq.~\ref{eq:refilling}) achieves the maximum at this point, therefore the steady-state fidelity (Eq.~\ref{eq:fidelity}) is also maximized at this point. 

Finally, we choose the maximum refilling rate set $W=\kappa$ in Fig.~\ref{suppfig:stable}(c) and sweep both the QQ sidebands rate $\Omega$ and resonator decay rate $\kappa$. The infidelity of the steady states is shown in contours. Larger $\Omega$ and $\kappa$ suppresses the infidelity efficiently. This indicates that our steady-state fidelity in the experiment is mainly limited by the sideband strengths. By increasing the QQ sidebands rate to above $2\pi\times\SI{10}{\mega\hertz}$, in simulation, it is possible to achieve stabilization fidelity above $98\%$.

\section{Other stabilization combinations}
\label{app:other}
Here we provide a list of states that can be stabilized with our protocol.

Case 1: Any two-qubit product states.
\begin{align}
\ket{\psi_{\phi_1, \phi_2}}=&(\cos(\phi_1/2)\ket{g}+\sin(\phi_1/2)\ket{e})\otimes\nonumber\\&(\cos(\phi_2/2)\ket{g}+\sin(\phi_2/2)\ket{e}). 
\label{appeq:case1}
\end{align}
This can be achieved by applying two detuned single qubit rabi drives on both $Q_1$ and $Q_2$ with rate $\{A_1, A_2\}$ and detunings $\{\delta_1, \delta_2\}$. The two-qubit rotating frame Hamiltonian becomes
\begin{align}
H_{p}=\begin{bmatrix}
               0 & A_2/2 & A_1/2 & 0 \\
               A_2/2 & \delta_2 & 0 & A_1/2 \\
               A_1/2 & 0 & \delta_1 & A_2/2 \\
               0 & A_1/2 & A_2/2 & \delta_1+\delta_2 
    \end{bmatrix}. \label{eq:H2} \\
\end{align}
It can be easily verified that the four eigenenergies $\{E_A<E_B<E_C<E_D\}$ satisfy the requirements $E_A+E_D=E_B+E_C$. Therefore, the lowest energy eigenstate can be efficiently stabilized by detuning two QR sideband frequencies. This is also a direct extension of the single-qubit stabilization scheme~\cite{Yao2017} to the two-qubit case.

Case 2: Dressed parity Bell states. The stabilized state set can be described by one continuous variable $\theta_1$
\begin{align}
\ket{\zeta_{\theta_1}}=&\cos(\theta_1/2)\ket{\Psi_{-}}+\sin(\theta_1/2)\ket{\Phi_{-}}.
\label{appeq:case2}
\end{align}
In this case, we apply on-resonant QQ blue and single qubit rabi drive on $Q_1$ with rate $\Omega$ and $A_{1}$ to dress the stabilized state's parity. The two-qubit Hamiltonian $H_{b}$ can be written as
\begin{align}
H_{b}=\begin{bmatrix}
               0 & 0 & A_1/2 & \Omega/2 \\
               0 & 0 & 0 & A_1/2 \\
               A_1/2 & 0 & 0 & 0 \\
               \Omega/2 & A_1/2 & 0 & 0 
    \end{bmatrix}. \label{eq:H33}
\end{align}
\begin{figure}[t]
    \centering
    \includegraphics[width=\columnwidth]{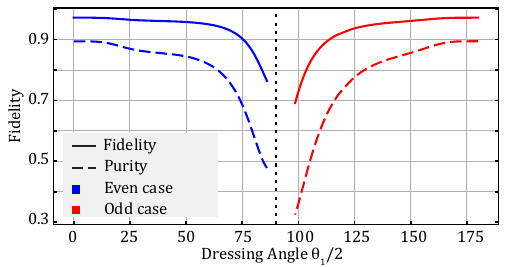}
    \caption{Stabilizing a one-dimensional set of entangled states. Blue and red lines represent separately using the QQ blue and QQ red sideband in the rotating frame simulation. Parameter used in the simulation: $\{\Omega, W_1, W_2, \kappa_1, \kappa_2\}/2\pi=\{5.0, 0.5, 0.5, 0.3, 0.33\}$ MHz. Qubit coherence time are $\{T_1, T_{\phi}\}=\{30, 30\} \mu$s.}
    \centering
    \label{suppfig:dressed_parity}
\end{figure}

One can verify that the four eigenenergies $E_A<E_B,<E_C<E_D$ of $H_{b}$ satisfy the requirement $E_A+E_D=E_B+E_C$. By choosing QR sidebands detunings as $E_B-E_A$ and $E_C-E_A$, the eigenstate $\ket{A}=\ket{\zeta_{\theta_1}}$ is stabilized, with the dressing angle being
\begin{align}
\theta_{1}=2\arctan\left(\frac{2A_1}{\Omega+\sqrt{4A_1^2+\Omega^2}}\right).
\end{align}

Similarly, we apply on-resonant QQ red sideband and single qubit rabi drive on $Q_1$ with rate $\Omega$ and $A_{1}$. The two-qubit Hamiltonian $H_{r}$ is 
\begin{align}
H_{r}=\begin{bmatrix}
               0 & 0 & A_1/2 & 0 \\
               0 & 0 & \Omega/2 & A_1/2 \\
               A_1/2 & \Omega/2 & 0 & 0 \\
               0 & A_1/2 & 0 & 0 
    \end{bmatrix}. \label{eq:H4}
\end{align}
The dressing angle $\theta_1$ under this case is 
\begin{align}
\theta_{1}=\pi-2\arctan\left(\frac{2A_1}{\Omega+\sqrt{4A_1^2+\Omega^2}}\right).
\end{align}

Steady-state fidelity is calculated through Qutip simulation for both cases, shown in Fig.~\ref{suppfig:dressed_parity}. By combining different QQ sideband colors, all dressing angles are stabilized with the scheme, except the small band region around $\theta_1=\pi$ where the effective transition rates provided by QR sidebands are close to $0$.

\begin{figure}[t]
    \centering
    \includegraphics[width=\columnwidth]{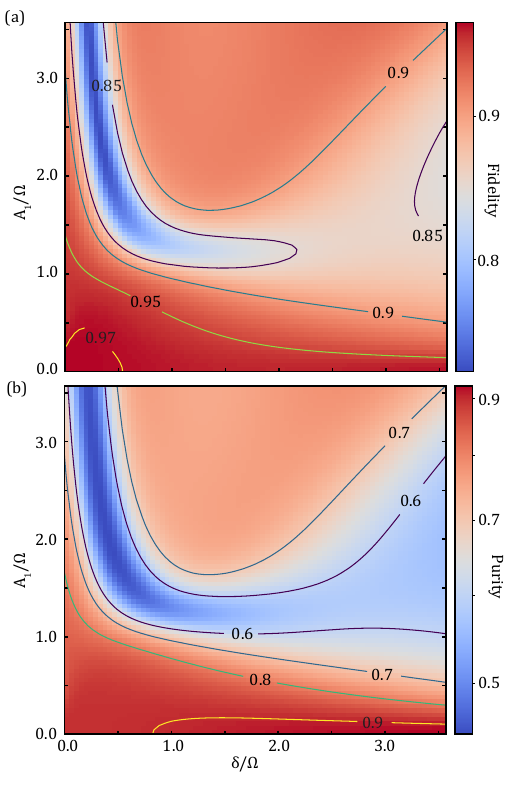}
    \caption{Stabilizing a 2D set of entangled states. Steady states' fidelity (a) and purity (b) are simulated in the rotating frame and plotted. $\frac{A_1}{\Omega}$ and $\frac{\delta}{\Omega}$ are separately two free variables that are swept to stabilize different states. Parameter used in the simulation: $\{\Omega, W_1, W_2, \kappa_1, \kappa_2\}/2\pi=\{5.0, 0.5, 0.5, 0.3, 0.33\}$ MHz. Qubit coherence time are$\{T_1, T_{\phi}\}=\{30, 30\} \mu$s.}
    \centering
    \label{suppfig:dressed_knob}
\end{figure}

Case 3: Rabi-dressed entangled states. This is a more general case where a two-dimensional set of entangled states is stabilized. We simultaneously apply a single qubit rabi drive on $Q_1$ with the rate $A_1$ and detuned QQ blue sideband with rate $\Omega$ and detuning $\delta$. The rotating frame Hamiltonian is 
\begin{align}
H_{g}=\begin{bmatrix}
               -\delta/2 & 0 & A_1/2 & \Omega/2 \\
               0 & \delta/2 & 0 & A_1/2 \\
               A_1/2 & 0 & -\delta/2 & 0 \\
               \Omega/2 & A_1/2 & 0 & \delta/2 
    \end{bmatrix}. \label{eq:H5} 
\end{align}
The four eigenenergies of the Hamiltonian can also be grouped into two pairs sharing the same sum. By appropriately choosing two QR sideband detunings, the lowest energy eigenstate is stabilized 
\begin{align}
\ket{\xi_{\delta, A_1}}=&E_{00}\ket{gg}+E_{01}\ket{ge}+E_{10}\ket{eg}-\ket{ee}, \nonumber \\
x =& \sqrt{4\delta^2A_1^2+4A_1^2\Omega^2+\Omega^4}, \nonumber \\
y =& \sqrt{\delta^2+2(2A_1^2+\Omega^2+x)}, \nonumber \\
E_{00} =& \frac{(\delta-y)(\delta^2+\Omega^2+x+\delta y)}{2\Omega(2A_1^2+\Omega^2+x)}, \nonumber \\
E_{01} = & \frac{A_1(\delta-y)}{2A_1^2+\Omega^2+x}, \nonumber \\
E_{10} = & -\frac{A_1(\delta^2+\Omega^2+x+\delta y)}{\Omega(2A_1^2+\Omega^2+x)}.
\label{appeq:case3}
\end{align}

For brevity, the stabilized state $\ket{\xi_{\delta, A_1}}$ is not normalized. The form of the state is determined by two independent variables $\frac{\delta}{\Omega}$ and $\frac{A_1}{\Omega}$. We sweep these two variables and plot the simulated fidelity and purity in Fig.~\ref{suppfig:dressed_knob}. This a general map covering all stabilized entangled states in the programmable operation: the vertical cut $\frac{\delta}{\Omega}=0$ represents the blue line in case 2 Dressed parity Bell states, the horizontal cut $\frac{A_1}{\Omega}=0$ represents the stabilized states shown in Fig.~\ref{fig:theory}(b), and the bottom left point $(\frac{\delta}{\Omega}, \frac{A_1}{\Omega})=(0,0)$ is the even parity bell state $\ket{\Psi_{-}}$. Using a modest sideband rate combination, most of the states on the plot can be stabilized with fidelity over $90\%$. Correspondingly, changing the QQ sideband color to red can stabilize another 2D set of entangled states which are dual to this case. All possible programmable stabilization operations can be chosen accordingly through the map provided here. States with both $E_{01}>1$ and $E_{10}>1$ cannot be stabilized under this case for instance.

\section{QR sideband colors and detunings}
\label{app:color}

Changing sideband colors and detuning frequency signs stabilize different states. This is important for stabilizing $\ket{\Psi_{\theta}}$ and $\ket{\Phi_{\theta}}$: at certain blending angles the steady state fidelity is low because of the small effective refilling rate $\Gamma_{t}$ (See Eq.~\ref{eq:refilling}). 

\begin{figure}[t]
    \centering
    \includegraphics[width=\columnwidth]{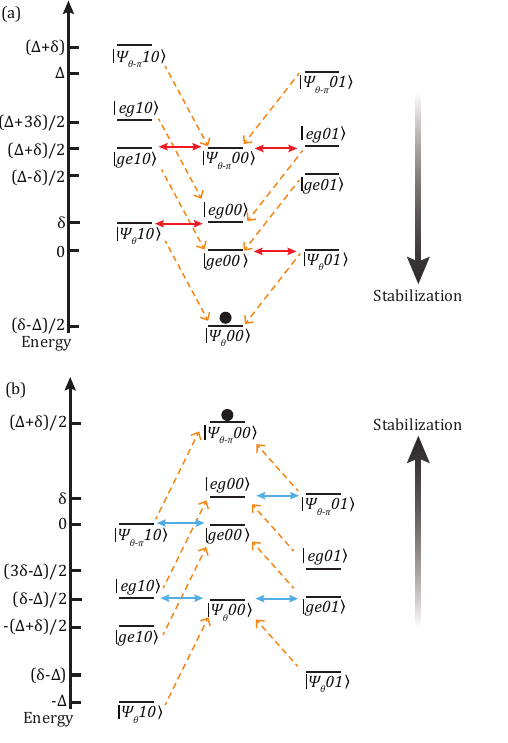}
    \caption{(a) Use both QR red sidebands to stabilize $\ket{\Psi_{\theta}}$. (b) Use opposite QR blue sideband detunings to stabilize $\ket{\Psi_{\theta-\pi}}$. The QQ sideband rates and detunings are separate $\Omega$ and $\delta$, and the QR sideband is detuned in frequency by $\frac{\Delta+\delta}{2}$, $\frac{\Delta-\delta}{2}$ in (a) and $-\frac{\Delta+\delta}{2}$, $-\frac{\Delta-\delta}{2}$ in (b). Here $\Delta=\sqrt{\Omega^2+\delta^2}$.}
    \centering
    \label{suppfig:color_detuning}
\end{figure}

As an explicit example, we consider the $\ket{\Psi_{\theta}}$ stabilization case. Instead of using two QR blue sidebands, we use two QR red sidebands with the same detuning and sideband rate. The rotating frame Hamiltonian now becomes
\begin{align}
    {H_{\rm color}} = & \frac{\Omega}{2}\left(a_{q1}a_{q2}+h.c.\right)+\delta a_{q1}^{\dag}a_{q1} \nonumber\\ 
    & +\frac{W_{1}}{2}\left(a_{q1}^{\dag}a_{r1}+h.c.\right)+\frac{W_{2}}{2}\left(a_{q2}^{\dag}a_{r2}+h.c.\right) \nonumber\\ 
    & +\frac{\Delta+\delta}{2}a_{r1}^{\dag}a_{r1}+\frac{\Delta-\delta}{2}a_{r2}^{\dag}a_{r2}.
    \label{eq:detuning_red}
\end{align}

Fig.~\ref{suppfig:color_detuning} (a) shows the level diagram, where the QR red sidebands now connect $\ket{ge00}$ to $\ket{\Psi_{\theta}01}$, which is different from the QR blue sidebands case where $\ket{ge00}$ and $\ket{\Psi_{\theta}10}$ are connected. The two-step refilling rate $\Gamma_{tc}$ for $\ket{ge00}\rightarrow\ket{\Psi_{\theta}00}$ is 
\begin{align}
\Gamma_{tc} & =\frac{W^2\sin^2\left(\theta/2\right)\kappa}{\kappa^2+W^2\sin^2\left(\theta/2\right)}.
\end{align}
The other three two-step transitions $\ket{eg00}\rightarrow\ket{\Psi_{\theta}00}$, $\ket{\Psi_{\theta-\pi}00}\rightarrow\ket{ge00}$, and $\ket{\Psi_{\theta-\pi}00}\rightarrow\ket{eg00}$ have the same refilling rate. Therefore, the steady-state fidelity for $\ket{\Psi_{\theta}}$ is
\begin{align}
\mathcal{F}_{\infty}=\left(\frac{\Gamma_t+\gamma\cos^2\left(\theta/2\right)}{\Gamma_t+\gamma}\right)^2.
\label{eq:fidelity_color}
\end{align}
One can thus choosing the QR sideband color for higher $F_{\infty}$. When the $F_{\infty}$ drops significantly near $\theta=\pi$ for $\Phi_{\theta}$ stabilization case, one can flip the QR sideband color for better performance.

We can also keep the same QR sideband color while choosing opposite QR sideband detunings. The Hamiltonian becomes
\begin{align}
    {H_{\rm opp}} = & \frac{\Omega}{2}\left(a_{q1}a_{q2}+h.c.\right)+\delta a_{q1}^{\dag}a_{q1} \nonumber\\ 
    & +\frac{W_{1}}{2}\left(a_{q1}^{\dag}a_{r1}+h.c.\right)+\frac{W_{2}}{2}\left(a_{q2}^{\dag}a_{r2}+h.c.\right) \nonumber\\ 
    & -\frac{\Delta+\delta}{2}a_{r1}^{\dag}a_{r1}-\frac{\Delta-\delta}{2}a_{r2}^{\dag}a_{r2}.
    \label{eq:detuning_opp}
\end{align}
The level diagram is shown in Fig.~\ref{suppfig:color_detuning} (b). All population flows to $\ket{\Psi_{\theta-\pi}}$, with the same refilling rate (Eq.~\ref{eq:refilling}) and steady-state fidelity (Eq.~\ref{eq:fidelity}) as the $\ket{\Psi_{\theta}}$ case.

\section{Measurement setup}

\begin{figure*}[b]
    \centering
    \includegraphics[width=2\columnwidth]{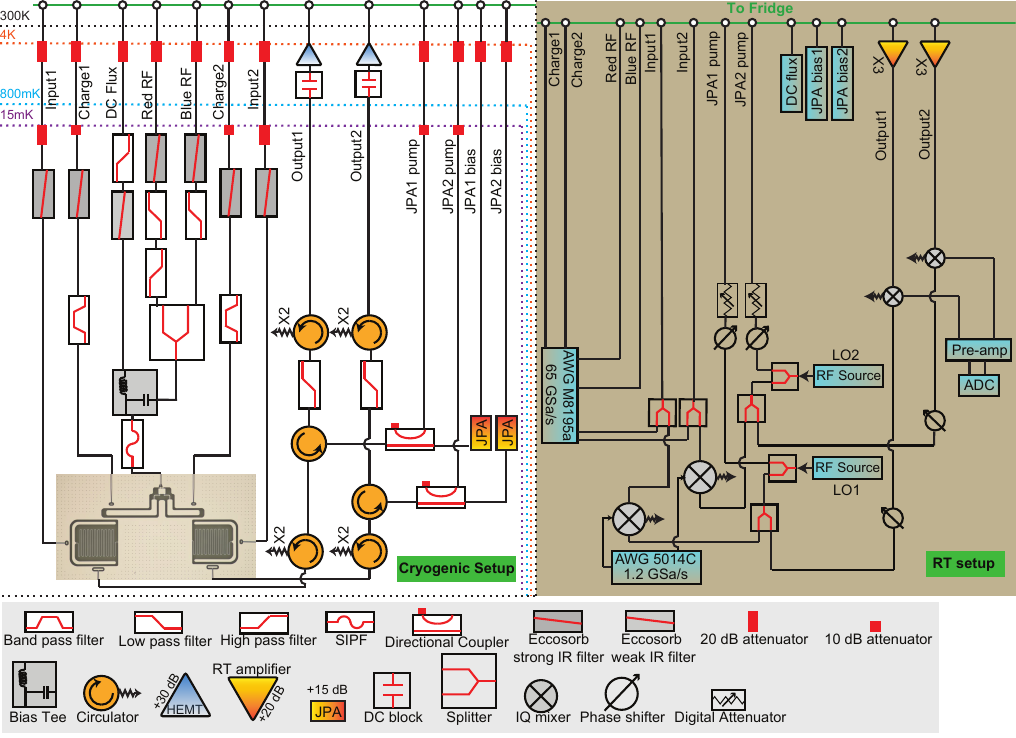}
    \caption{Detailed measurement setup.}
    \centering
    \label{suppfig:setup}
\end{figure*}

Figure~\ref{suppfig:setup} illustrates the measurement setup used in the experiment. Both single qubit, QR sidebands, and QQ sidebands signals are generated with a 4-channel AWG (Keysight M8195 65 Gsa/s, 16 Gsa/s per channel) to maintain phase-locking. DC flux bias for the coupler and two Josephson Parametric Amplifiers (JPAa) are generated with three current sources (Yokogawa GS200). The bandpass filters on both charge lines are chosen such that the stop band covers both readout and qubits' frequencies. The flux drive is delivered using three separate coaxial cables for DC bias, red sideband and blue sideband frequencies. The RF lines are merged using a combiner which is then added to the DC bias using a bias tee. The Stepped impedance Purcell filter (SIPF) inserted in the flux line is a home-made filter with a stop band between $\SI{2}{\giga\hertz}$ and $\SI{5.5}{\giga\hertz}$ blocking any qubit and resonator signals. Each transmitted signal is first amplified by a JPA with a $+15$ dB gain, followed by a HEMT amplifier and three room-temperature amplifiers. The final signal is pre-amplified after demodulation and digitized with an Alazar ATS 9870 (1GSa/s) card. 

\bibliography{main}

\begin{thebibliography}{27}%
\makeatletter
\providecommand \@ifxundefined [1]{%
 \@ifx{#1\undefined}
}%
\providecommand \@ifnum [1]{%
 \ifnum #1\expandafter \@firstoftwo
 \else \expandafter \@secondoftwo
 \fi
}%
\providecommand \@ifx [1]{%
 \ifx #1\expandafter \@firstoftwo
 \else \expandafter \@secondoftwo
 \fi
}%
\providecommand \natexlab [1]{#1}%
\providecommand \enquote  [1]{``#1''}%
\providecommand \bibnamefont  [1]{#1}%
\providecommand \bibfnamefont [1]{#1}%
\providecommand \citenamefont [1]{#1}%
\providecommand \href@noop [0]{\@secondoftwo}%
\providecommand \href [0]{\begingroup \@sanitize@url \@href}%
\providecommand \@href[1]{\@@startlink{#1}\@@href}%
\providecommand \@@href[1]{\endgroup#1\@@endlink}%
\providecommand \@sanitize@url [0]{\catcode `\\12\catcode `\$12\catcode
  `\&12\catcode `\#12\catcode `\^12\catcode `\_12\catcode `\%12\relax}%
\providecommand \@@startlink[1]{}%
\providecommand \@@endlink[0]{}%
\providecommand \url  [0]{\begingroup\@sanitize@url \@url }%
\providecommand \@url [1]{\endgroup\@href {#1}{\urlprefix }}%
\providecommand \urlprefix  [0]{URL }%
\providecommand \Eprint [0]{\href }%
\providecommand \doibase [0]{https://doi.org/}%
\providecommand \selectlanguage [0]{\@gobble}%
\providecommand \bibinfo  [0]{\@secondoftwo}%
\providecommand \bibfield  [0]{\@secondoftwo}%
\providecommand \translation [1]{[#1]}%
\providecommand \BibitemOpen [0]{}%
\providecommand \bibitemStop [0]{}%
\providecommand \bibitemNoStop [0]{.\EOS\space}%
\providecommand \EOS [0]{\spacefactor3000\relax}%
\providecommand \BibitemShut  [1]{\csname bibitem#1\endcsname}%
\let\auto@bib@innerbib\@empty
\bibitem [{\citenamefont {Arute}\ \emph {et~al.}(2019)\citenamefont {Arute},
  \citenamefont {Arya}, \citenamefont {Babbush}, \citenamefont {Bacon},
  \citenamefont {Bardin}, \citenamefont {Barends}, \citenamefont {Biswas},
  \citenamefont {Boixo}, \citenamefont {Brandao}, \citenamefont {Buell},
  \citenamefont {Burkett}, \citenamefont {Chen}, \citenamefont {Chen},
  \citenamefont {Chiaro}, \citenamefont {Collins}, \citenamefont {Courtney},
  \citenamefont {Dunsworth}, \citenamefont {Farhi}, \citenamefont {Foxen},
  \citenamefont {Fowler}, \citenamefont {Gidney}, \citenamefont {Giustina},
  \citenamefont {Graff}, \citenamefont {Guerin}, \citenamefont {Habegger},
  \citenamefont {Harrigan}, \citenamefont {Hartmann}, \citenamefont {Ho},
  \citenamefont {Hoffmann}, \citenamefont {Huang}, \citenamefont {Humble},
  \citenamefont {Isakov}, \citenamefont {Jeffrey}, \citenamefont {Jiang},
  \citenamefont {Kafri}, \citenamefont {Kechedzhi}, \citenamefont {Kelly},
  \citenamefont {Klimov}, \citenamefont {Knysh}, \citenamefont {Korotkov},
  \citenamefont {Kostritsa}, \citenamefont {Landhuis}, \citenamefont
  {Lindmark}, \citenamefont {Lucero}, \citenamefont {Lyakh}, \citenamefont
  {Mandr{\`a}}, \citenamefont {McClean}, \citenamefont {McEwen}, \citenamefont
  {Megrant}, \citenamefont {Mi}, \citenamefont {Michielsen}, \citenamefont
  {Mohseni}, \citenamefont {Mutus}, \citenamefont {Naaman}, \citenamefont
  {Neeley}, \citenamefont {Neill}, \citenamefont {Niu}, \citenamefont {Ostby},
  \citenamefont {Petukhov}, \citenamefont {Platt}, \citenamefont {Quintana},
  \citenamefont {Rieffel}, \citenamefont {Roushan}, \citenamefont {Rubin},
  \citenamefont {Sank}, \citenamefont {Satzinger}, \citenamefont {Smelyanskiy},
  \citenamefont {Sung}, \citenamefont {Trevithick}, \citenamefont
  {Vainsencher}, \citenamefont {Villalonga}, \citenamefont {White},
  \citenamefont {Yao}, \citenamefont {Yeh}, \citenamefont {Zalcman},
  \citenamefont {Neven},\ and\ \citenamefont {Martinis}}]{Arute2019}%
  \BibitemOpen
  \bibfield  {author} {\bibinfo {author} {\bibfnamefont {F.}~\bibnamefont
  {Arute}}, \bibinfo {author} {\bibfnamefont {K.}~\bibnamefont {Arya}},
  \bibinfo {author} {\bibfnamefont {R.}~\bibnamefont {Babbush}}, \bibinfo
  {author} {\bibfnamefont {D.}~\bibnamefont {Bacon}}, \bibinfo {author}
  {\bibfnamefont {J.~C.}\ \bibnamefont {Bardin}}, \bibinfo {author}
  {\bibfnamefont {R.}~\bibnamefont {Barends}}, \bibinfo {author} {\bibfnamefont
  {R.}~\bibnamefont {Biswas}}, \bibinfo {author} {\bibfnamefont
  {S.}~\bibnamefont {Boixo}}, \bibinfo {author} {\bibfnamefont {F.~G. S.~L.}\
  \bibnamefont {Brandao}}, \bibinfo {author} {\bibfnamefont {D.~A.}\
  \bibnamefont {Buell}}, \bibinfo {author} {\bibfnamefont {B.}~\bibnamefont
  {Burkett}}, \bibinfo {author} {\bibfnamefont {Y.}~\bibnamefont {Chen}},
  \bibinfo {author} {\bibfnamefont {Z.}~\bibnamefont {Chen}}, \bibinfo {author}
  {\bibfnamefont {B.}~\bibnamefont {Chiaro}}, \bibinfo {author} {\bibfnamefont
  {R.}~\bibnamefont {Collins}}, \bibinfo {author} {\bibfnamefont
  {W.}~\bibnamefont {Courtney}}, \bibinfo {author} {\bibfnamefont
  {A.}~\bibnamefont {Dunsworth}}, \bibinfo {author} {\bibfnamefont
  {E.}~\bibnamefont {Farhi}}, \bibinfo {author} {\bibfnamefont
  {B.}~\bibnamefont {Foxen}}, \bibinfo {author} {\bibfnamefont
  {A.}~\bibnamefont {Fowler}}, \bibinfo {author} {\bibfnamefont
  {C.}~\bibnamefont {Gidney}}, \bibinfo {author} {\bibfnamefont
  {M.}~\bibnamefont {Giustina}}, \bibinfo {author} {\bibfnamefont
  {R.}~\bibnamefont {Graff}}, \bibinfo {author} {\bibfnamefont
  {K.}~\bibnamefont {Guerin}}, \bibinfo {author} {\bibfnamefont
  {S.}~\bibnamefont {Habegger}}, \bibinfo {author} {\bibfnamefont {M.~P.}\
  \bibnamefont {Harrigan}}, \bibinfo {author} {\bibfnamefont {M.~J.}\
  \bibnamefont {Hartmann}}, \bibinfo {author} {\bibfnamefont {A.}~\bibnamefont
  {Ho}}, \bibinfo {author} {\bibfnamefont {M.}~\bibnamefont {Hoffmann}},
  \bibinfo {author} {\bibfnamefont {T.}~\bibnamefont {Huang}}, \bibinfo
  {author} {\bibfnamefont {T.~S.}\ \bibnamefont {Humble}}, \bibinfo {author}
  {\bibfnamefont {S.~V.}\ \bibnamefont {Isakov}}, \bibinfo {author}
  {\bibfnamefont {E.}~\bibnamefont {Jeffrey}}, \bibinfo {author} {\bibfnamefont
  {Z.}~\bibnamefont {Jiang}}, \bibinfo {author} {\bibfnamefont
  {D.}~\bibnamefont {Kafri}}, \bibinfo {author} {\bibfnamefont
  {K.}~\bibnamefont {Kechedzhi}}, \bibinfo {author} {\bibfnamefont
  {J.}~\bibnamefont {Kelly}}, \bibinfo {author} {\bibfnamefont {P.~V.}\
  \bibnamefont {Klimov}}, \bibinfo {author} {\bibfnamefont {S.}~\bibnamefont
  {Knysh}}, \bibinfo {author} {\bibfnamefont {A.}~\bibnamefont {Korotkov}},
  \bibinfo {author} {\bibfnamefont {F.}~\bibnamefont {Kostritsa}}, \bibinfo
  {author} {\bibfnamefont {D.}~\bibnamefont {Landhuis}}, \bibinfo {author}
  {\bibfnamefont {M.}~\bibnamefont {Lindmark}}, \bibinfo {author}
  {\bibfnamefont {E.}~\bibnamefont {Lucero}}, \bibinfo {author} {\bibfnamefont
  {D.}~\bibnamefont {Lyakh}}, \bibinfo {author} {\bibfnamefont
  {S.}~\bibnamefont {Mandr{\`a}}}, \bibinfo {author} {\bibfnamefont {J.~R.}\
  \bibnamefont {McClean}}, \bibinfo {author} {\bibfnamefont {M.}~\bibnamefont
  {McEwen}}, \bibinfo {author} {\bibfnamefont {A.}~\bibnamefont {Megrant}},
  \bibinfo {author} {\bibfnamefont {X.}~\bibnamefont {Mi}}, \bibinfo {author}
  {\bibfnamefont {K.}~\bibnamefont {Michielsen}}, \bibinfo {author}
  {\bibfnamefont {M.}~\bibnamefont {Mohseni}}, \bibinfo {author} {\bibfnamefont
  {J.}~\bibnamefont {Mutus}}, \bibinfo {author} {\bibfnamefont
  {O.}~\bibnamefont {Naaman}}, \bibinfo {author} {\bibfnamefont
  {M.}~\bibnamefont {Neeley}}, \bibinfo {author} {\bibfnamefont
  {C.}~\bibnamefont {Neill}}, \bibinfo {author} {\bibfnamefont {M.~Y.}\
  \bibnamefont {Niu}}, \bibinfo {author} {\bibfnamefont {E.}~\bibnamefont
  {Ostby}}, \bibinfo {author} {\bibfnamefont {A.}~\bibnamefont {Petukhov}},
  \bibinfo {author} {\bibfnamefont {J.~C.}\ \bibnamefont {Platt}}, \bibinfo
  {author} {\bibfnamefont {C.}~\bibnamefont {Quintana}}, \bibinfo {author}
  {\bibfnamefont {E.~G.}\ \bibnamefont {Rieffel}}, \bibinfo {author}
  {\bibfnamefont {P.}~\bibnamefont {Roushan}}, \bibinfo {author} {\bibfnamefont
  {N.~C.}\ \bibnamefont {Rubin}}, \bibinfo {author} {\bibfnamefont
  {D.}~\bibnamefont {Sank}}, \bibinfo {author} {\bibfnamefont {K.~J.}\
  \bibnamefont {Satzinger}}, \bibinfo {author} {\bibfnamefont {V.}~\bibnamefont
  {Smelyanskiy}}, \bibinfo {author} {\bibfnamefont {K.~J.}\ \bibnamefont
  {Sung}}, \bibinfo {author} {\bibfnamefont {M.~D.}\ \bibnamefont
  {Trevithick}}, \bibinfo {author} {\bibfnamefont {A.}~\bibnamefont
  {Vainsencher}}, \bibinfo {author} {\bibfnamefont {B.}~\bibnamefont
  {Villalonga}}, \bibinfo {author} {\bibfnamefont {T.}~\bibnamefont {White}},
  \bibinfo {author} {\bibfnamefont {Z.~J.}\ \bibnamefont {Yao}}, \bibinfo
  {author} {\bibfnamefont {P.}~\bibnamefont {Yeh}}, \bibinfo {author}
  {\bibfnamefont {A.}~\bibnamefont {Zalcman}}, \bibinfo {author} {\bibfnamefont
  {H.}~\bibnamefont {Neven}},\ and\ \bibinfo {author} {\bibfnamefont {J.~M.}\
  \bibnamefont {Martinis}},\ }\bibfield  {title} {\bibinfo {title} {Quantum
  supremacy using a programmable superconducting processor},\ }\href
  {https://doi.org/10.1038/s41586-019-1666-5} {\bibfield  {journal} {\bibinfo
  {journal} {Nature}\ }\textbf {\bibinfo {volume} {574}},\ \bibinfo {pages}
  {505} (\bibinfo {year} {2019})}\BibitemShut {NoStop}%
\bibitem [{\citenamefont {Wu}\ \emph {et~al.}(2021)\citenamefont {Wu},
  \citenamefont {Bao}, \citenamefont {Cao}, \citenamefont {Chen}, \citenamefont
  {Chen}, \citenamefont {Chen}, \citenamefont {Chung}, \citenamefont {Deng},
  \citenamefont {Du}, \citenamefont {Fan}, \citenamefont {Gong}, \citenamefont
  {Guo}, \citenamefont {Guo}, \citenamefont {Guo}, \citenamefont {Han},
  \citenamefont {Hong}, \citenamefont {Huang}, \citenamefont {Huo},
  \citenamefont {Li}, \citenamefont {Li}, \citenamefont {Li}, \citenamefont
  {Li}, \citenamefont {Liang}, \citenamefont {Lin}, \citenamefont {Lin},
  \citenamefont {Qian}, \citenamefont {Qiao}, \citenamefont {Rong},
  \citenamefont {Su}, \citenamefont {Sun}, \citenamefont {Wang}, \citenamefont
  {Wang}, \citenamefont {Wu}, \citenamefont {Xu}, \citenamefont {Yan},
  \citenamefont {Yang}, \citenamefont {Yang}, \citenamefont {Ye}, \citenamefont
  {Yin}, \citenamefont {Ying}, \citenamefont {Yu}, \citenamefont {Zha},
  \citenamefont {Zhang}, \citenamefont {Zhang}, \citenamefont {Zhang},
  \citenamefont {Zhang}, \citenamefont {Zhao}, \citenamefont {Zhao},
  \citenamefont {Zhou}, \citenamefont {Zhu}, \citenamefont {Lu}, \citenamefont
  {Peng}, \citenamefont {Zhu},\ and\ \citenamefont {Pan}}]{Wu2021}%
  \BibitemOpen
  \bibfield  {author} {\bibinfo {author} {\bibfnamefont {Y.}~\bibnamefont
  {Wu}}, \bibinfo {author} {\bibfnamefont {W.-S.}\ \bibnamefont {Bao}},
  \bibinfo {author} {\bibfnamefont {S.}~\bibnamefont {Cao}}, \bibinfo {author}
  {\bibfnamefont {F.}~\bibnamefont {Chen}}, \bibinfo {author} {\bibfnamefont
  {M.-C.}\ \bibnamefont {Chen}}, \bibinfo {author} {\bibfnamefont
  {X.}~\bibnamefont {Chen}}, \bibinfo {author} {\bibfnamefont {T.-H.}\
  \bibnamefont {Chung}}, \bibinfo {author} {\bibfnamefont {H.}~\bibnamefont
  {Deng}}, \bibinfo {author} {\bibfnamefont {Y.}~\bibnamefont {Du}}, \bibinfo
  {author} {\bibfnamefont {D.}~\bibnamefont {Fan}}, \bibinfo {author}
  {\bibfnamefont {M.}~\bibnamefont {Gong}}, \bibinfo {author} {\bibfnamefont
  {C.}~\bibnamefont {Guo}}, \bibinfo {author} {\bibfnamefont {C.}~\bibnamefont
  {Guo}}, \bibinfo {author} {\bibfnamefont {S.}~\bibnamefont {Guo}}, \bibinfo
  {author} {\bibfnamefont {L.}~\bibnamefont {Han}}, \bibinfo {author}
  {\bibfnamefont {L.}~\bibnamefont {Hong}}, \bibinfo {author} {\bibfnamefont
  {H.-L.}\ \bibnamefont {Huang}}, \bibinfo {author} {\bibfnamefont {Y.-H.}\
  \bibnamefont {Huo}}, \bibinfo {author} {\bibfnamefont {L.}~\bibnamefont
  {Li}}, \bibinfo {author} {\bibfnamefont {N.}~\bibnamefont {Li}}, \bibinfo
  {author} {\bibfnamefont {S.}~\bibnamefont {Li}}, \bibinfo {author}
  {\bibfnamefont {Y.}~\bibnamefont {Li}}, \bibinfo {author} {\bibfnamefont
  {F.}~\bibnamefont {Liang}}, \bibinfo {author} {\bibfnamefont
  {C.}~\bibnamefont {Lin}}, \bibinfo {author} {\bibfnamefont {J.}~\bibnamefont
  {Lin}}, \bibinfo {author} {\bibfnamefont {H.}~\bibnamefont {Qian}}, \bibinfo
  {author} {\bibfnamefont {D.}~\bibnamefont {Qiao}}, \bibinfo {author}
  {\bibfnamefont {H.}~\bibnamefont {Rong}}, \bibinfo {author} {\bibfnamefont
  {H.}~\bibnamefont {Su}}, \bibinfo {author} {\bibfnamefont {L.}~\bibnamefont
  {Sun}}, \bibinfo {author} {\bibfnamefont {L.}~\bibnamefont {Wang}}, \bibinfo
  {author} {\bibfnamefont {S.}~\bibnamefont {Wang}}, \bibinfo {author}
  {\bibfnamefont {D.}~\bibnamefont {Wu}}, \bibinfo {author} {\bibfnamefont
  {Y.}~\bibnamefont {Xu}}, \bibinfo {author} {\bibfnamefont {K.}~\bibnamefont
  {Yan}}, \bibinfo {author} {\bibfnamefont {W.}~\bibnamefont {Yang}}, \bibinfo
  {author} {\bibfnamefont {Y.}~\bibnamefont {Yang}}, \bibinfo {author}
  {\bibfnamefont {Y.}~\bibnamefont {Ye}}, \bibinfo {author} {\bibfnamefont
  {J.}~\bibnamefont {Yin}}, \bibinfo {author} {\bibfnamefont {C.}~\bibnamefont
  {Ying}}, \bibinfo {author} {\bibfnamefont {J.}~\bibnamefont {Yu}}, \bibinfo
  {author} {\bibfnamefont {C.}~\bibnamefont {Zha}}, \bibinfo {author}
  {\bibfnamefont {C.}~\bibnamefont {Zhang}}, \bibinfo {author} {\bibfnamefont
  {H.}~\bibnamefont {Zhang}}, \bibinfo {author} {\bibfnamefont
  {K.}~\bibnamefont {Zhang}}, \bibinfo {author} {\bibfnamefont
  {Y.}~\bibnamefont {Zhang}}, \bibinfo {author} {\bibfnamefont
  {H.}~\bibnamefont {Zhao}}, \bibinfo {author} {\bibfnamefont {Y.}~\bibnamefont
  {Zhao}}, \bibinfo {author} {\bibfnamefont {L.}~\bibnamefont {Zhou}}, \bibinfo
  {author} {\bibfnamefont {Q.}~\bibnamefont {Zhu}}, \bibinfo {author}
  {\bibfnamefont {C.-Y.}\ \bibnamefont {Lu}}, \bibinfo {author} {\bibfnamefont
  {C.-Z.}\ \bibnamefont {Peng}}, \bibinfo {author} {\bibfnamefont
  {X.}~\bibnamefont {Zhu}},\ and\ \bibinfo {author} {\bibfnamefont {J.-W.}\
  \bibnamefont {Pan}},\ }\bibfield  {title} {\bibinfo {title} {Strong quantum
  computational advantage using a superconducting quantum processor},\ }\href
  {https://doi.org/10.1103/PhysRevLett.127.180501} {\bibfield  {journal}
  {\bibinfo  {journal} {Phys. Rev. Lett.}\ }\textbf {\bibinfo {volume} {127}},\
  \bibinfo {pages} {180501} (\bibinfo {year} {2021})}\BibitemShut {NoStop}%
\bibitem [{\citenamefont {Kapit}\ \emph {et~al.}(2020)\citenamefont {Kapit},
  \citenamefont {Roushan}, \citenamefont {Neill}, \citenamefont {Boixo},\ and\
  \citenamefont {Smelyanskiy}}]{PhysRevResearch.2.043042}%
  \BibitemOpen
  \bibfield  {author} {\bibinfo {author} {\bibfnamefont {E.}~\bibnamefont
  {Kapit}}, \bibinfo {author} {\bibfnamefont {P.}~\bibnamefont {Roushan}},
  \bibinfo {author} {\bibfnamefont {C.}~\bibnamefont {Neill}}, \bibinfo
  {author} {\bibfnamefont {S.}~\bibnamefont {Boixo}},\ and\ \bibinfo {author}
  {\bibfnamefont {V.}~\bibnamefont {Smelyanskiy}},\ }\bibfield  {title}
  {\bibinfo {title} {Entanglement and complexity of interacting qubits subject
  to asymmetric noise},\ }\href
  {https://doi.org/10.1103/PhysRevResearch.2.043042} {\bibfield  {journal}
  {\bibinfo  {journal} {Phys. Rev. Res.}\ }\textbf {\bibinfo {volume} {2}},\
  \bibinfo {pages} {043042} (\bibinfo {year} {2020})}\BibitemShut {NoStop}%
\bibitem [{\citenamefont {Verstraete}\ \emph {et~al.}(2009)\citenamefont
  {Verstraete}, \citenamefont {Wolf},\ and\ \citenamefont
  {Ignacio~Cirac}}]{Verstraete2009}%
  \BibitemOpen
  \bibfield  {author} {\bibinfo {author} {\bibfnamefont {F.}~\bibnamefont
  {Verstraete}}, \bibinfo {author} {\bibfnamefont {M.~M.}\ \bibnamefont
  {Wolf}},\ and\ \bibinfo {author} {\bibfnamefont {J.}~\bibnamefont
  {Ignacio~Cirac}},\ }\bibfield  {title} {\bibinfo {title} {Quantum computation
  and quantum-state engineering driven by dissipation},\ }\href
  {https://doi.org/10.1038/nphys1342} {\bibfield  {journal} {\bibinfo
  {journal} {Nature Physics}\ }\textbf {\bibinfo {volume} {5}},\ \bibinfo
  {pages} {633} (\bibinfo {year} {2009})}\BibitemShut {NoStop}%
\bibitem [{\citenamefont {Kapit}(2016)}]{VSLQ2015}%
  \BibitemOpen
  \bibfield  {author} {\bibinfo {author} {\bibfnamefont {E.}~\bibnamefont
  {Kapit}},\ }\bibfield  {title} {\bibinfo {title} {Hardware-efficient and
  fully autonomous quantum error correction in superconducting circuits},\
  }\href {https://doi.org/10.1103/PhysRevLett.116.150501} {\bibfield  {journal}
  {\bibinfo  {journal} {Phys. Rev. Lett.}\ }\textbf {\bibinfo {volume} {116}},\
  \bibinfo {pages} {150501} (\bibinfo {year} {2016})}\BibitemShut {NoStop}%
\bibitem [{\citenamefont {Ma}\ \emph {et~al.}(2020)\citenamefont {Ma},
  \citenamefont {Xu}, \citenamefont {Mu}, \citenamefont {Cai}, \citenamefont
  {Hu}, \citenamefont {Wang}, \citenamefont {Pan}, \citenamefont {Wang},
  \citenamefont {Song}, \citenamefont {Zou},\ and\ \citenamefont
  {Sun}}]{Ma2020}%
  \BibitemOpen
  \bibfield  {author} {\bibinfo {author} {\bibfnamefont {Y.}~\bibnamefont
  {Ma}}, \bibinfo {author} {\bibfnamefont {Y.}~\bibnamefont {Xu}}, \bibinfo
  {author} {\bibfnamefont {X.}~\bibnamefont {Mu}}, \bibinfo {author}
  {\bibfnamefont {W.}~\bibnamefont {Cai}}, \bibinfo {author} {\bibfnamefont
  {L.}~\bibnamefont {Hu}}, \bibinfo {author} {\bibfnamefont {W.}~\bibnamefont
  {Wang}}, \bibinfo {author} {\bibfnamefont {X.}~\bibnamefont {Pan}}, \bibinfo
  {author} {\bibfnamefont {H.}~\bibnamefont {Wang}}, \bibinfo {author}
  {\bibfnamefont {Y.~P.}\ \bibnamefont {Song}}, \bibinfo {author}
  {\bibfnamefont {C.-L.}\ \bibnamefont {Zou}},\ and\ \bibinfo {author}
  {\bibfnamefont {L.}~\bibnamefont {Sun}},\ }\bibfield  {title} {\bibinfo
  {title} {Error-transparent operations on a logical qubit protected by quantum
  error correction},\ }\href {https://doi.org/10.1038/s41567-020-0893-x}
  {\bibfield  {journal} {\bibinfo  {journal} {Nature Physics}\ }\textbf
  {\bibinfo {volume} {16}},\ \bibinfo {pages} {827} (\bibinfo {year}
  {2020})}\BibitemShut {NoStop}%
\bibitem [{\citenamefont {Gertler}\ \emph {et~al.}(2021)\citenamefont
  {Gertler}, \citenamefont {Baker}, \citenamefont {Li}, \citenamefont {Shirol},
  \citenamefont {Koch},\ and\ \citenamefont {Wang}}]{Gertler2021}%
  \BibitemOpen
  \bibfield  {author} {\bibinfo {author} {\bibfnamefont {J.~M.}\ \bibnamefont
  {Gertler}}, \bibinfo {author} {\bibfnamefont {B.}~\bibnamefont {Baker}},
  \bibinfo {author} {\bibfnamefont {J.}~\bibnamefont {Li}}, \bibinfo {author}
  {\bibfnamefont {S.}~\bibnamefont {Shirol}}, \bibinfo {author} {\bibfnamefont
  {J.}~\bibnamefont {Koch}},\ and\ \bibinfo {author} {\bibfnamefont
  {C.}~\bibnamefont {Wang}},\ }\bibfield  {title} {\bibinfo {title} {Protecting
  a bosonic qubit with autonomous quantum error correction},\ }\href
  {https://doi.org/10.1038/s41586-021-03257-0} {\bibfield  {journal} {\bibinfo
  {journal} {Nature}\ }\textbf {\bibinfo {volume} {590}},\ \bibinfo {pages}
  {243} (\bibinfo {year} {2021})}\BibitemShut {NoStop}%
\bibitem [{\citenamefont {Li}\ \emph {et~al.}(2023{\natexlab{a}})\citenamefont
  {Li}, \citenamefont {Roy}, \citenamefont {Perez}, \citenamefont {Lee},
  \citenamefont {Kapit},\ and\ \citenamefont {Schuster}}]{li2023autonomous}%
  \BibitemOpen
  \bibfield  {author} {\bibinfo {author} {\bibfnamefont {Z.}~\bibnamefont
  {Li}}, \bibinfo {author} {\bibfnamefont {T.}~\bibnamefont {Roy}}, \bibinfo
  {author} {\bibfnamefont {D.~R.}\ \bibnamefont {Perez}}, \bibinfo {author}
  {\bibfnamefont {K.-H.}\ \bibnamefont {Lee}}, \bibinfo {author} {\bibfnamefont
  {E.}~\bibnamefont {Kapit}},\ and\ \bibinfo {author} {\bibfnamefont {D.~I.}\
  \bibnamefont {Schuster}},\ }\href@noop {} {\bibinfo {title} {Autonomous error
  correction of a single logical qubit using two transmons}} (\bibinfo {year}
  {2023}{\natexlab{a}}),\ \Eprint {https://arxiv.org/abs/2302.06707}
  {arXiv:2302.06707 [quant-ph]} \BibitemShut {NoStop}%
\bibitem [{\citenamefont {Li}\ \emph {et~al.}(2023{\natexlab{b}})\citenamefont
  {Li}, \citenamefont {Roy}, \citenamefont {Pérez}, \citenamefont {Schuster},\
  and\ \citenamefont {Kapit}}]{li2023hardware}%
  \BibitemOpen
  \bibfield  {author} {\bibinfo {author} {\bibfnamefont {Z.}~\bibnamefont
  {Li}}, \bibinfo {author} {\bibfnamefont {T.}~\bibnamefont {Roy}}, \bibinfo
  {author} {\bibfnamefont {D.~R.}\ \bibnamefont {Pérez}}, \bibinfo {author}
  {\bibfnamefont {D.~I.}\ \bibnamefont {Schuster}},\ and\ \bibinfo {author}
  {\bibfnamefont {E.}~\bibnamefont {Kapit}},\ }\href@noop {} {\bibinfo {title}
  {Hardware efficient autonomous error correction with linear couplers in
  superconducting circuits}} (\bibinfo {year} {2023}{\natexlab{b}}),\ \Eprint
  {https://arxiv.org/abs/2303.01110} {arXiv:2303.01110 [quant-ph]} \BibitemShut
  {NoStop}%
\bibitem [{\citenamefont {Shankar}\ \emph {et~al.}(2013)\citenamefont
  {Shankar}, \citenamefont {Hatridge}, \citenamefont {Leghtas}, \citenamefont
  {Sliwa}, \citenamefont {Narla}, \citenamefont {Vool}, \citenamefont {Girvin},
  \citenamefont {Frunzio}, \citenamefont {Mirrahimi},\ and\ \citenamefont
  {Devoret}}]{Shankar2013}%
  \BibitemOpen
  \bibfield  {author} {\bibinfo {author} {\bibfnamefont {S.}~\bibnamefont
  {Shankar}}, \bibinfo {author} {\bibfnamefont {M.}~\bibnamefont {Hatridge}},
  \bibinfo {author} {\bibfnamefont {Z.}~\bibnamefont {Leghtas}}, \bibinfo
  {author} {\bibfnamefont {K.~M.}\ \bibnamefont {Sliwa}}, \bibinfo {author}
  {\bibfnamefont {A.}~\bibnamefont {Narla}}, \bibinfo {author} {\bibfnamefont
  {U.}~\bibnamefont {Vool}}, \bibinfo {author} {\bibfnamefont {S.~M.}\
  \bibnamefont {Girvin}}, \bibinfo {author} {\bibfnamefont {L.}~\bibnamefont
  {Frunzio}}, \bibinfo {author} {\bibfnamefont {M.}~\bibnamefont {Mirrahimi}},\
  and\ \bibinfo {author} {\bibfnamefont {M.~H.}\ \bibnamefont {Devoret}},\
  }\bibfield  {title} {\bibinfo {title} {Autonomously stabilized entanglement
  between two superconducting quantum bits},\ }\href
  {https://doi.org/10.1038/nature12802} {\bibfield  {journal} {\bibinfo
  {journal} {Nature}\ }\textbf {\bibinfo {volume} {504}},\ \bibinfo {pages}
  {419} (\bibinfo {year} {2013})}\BibitemShut {NoStop}%
\bibitem [{\citenamefont {Kimchi-Schwartz}\ \emph {et~al.}(2016)\citenamefont
  {Kimchi-Schwartz}, \citenamefont {Martin}, \citenamefont {Flurin},
  \citenamefont {Aron}, \citenamefont {Kulkarni}, \citenamefont {Tureci},\ and\
  \citenamefont {Siddiqi}}]{siddiqi2016}%
  \BibitemOpen
  \bibfield  {author} {\bibinfo {author} {\bibfnamefont {M.~E.}\ \bibnamefont
  {Kimchi-Schwartz}}, \bibinfo {author} {\bibfnamefont {L.}~\bibnamefont
  {Martin}}, \bibinfo {author} {\bibfnamefont {E.}~\bibnamefont {Flurin}},
  \bibinfo {author} {\bibfnamefont {C.}~\bibnamefont {Aron}}, \bibinfo {author}
  {\bibfnamefont {M.}~\bibnamefont {Kulkarni}}, \bibinfo {author}
  {\bibfnamefont {H.~E.}\ \bibnamefont {Tureci}},\ and\ \bibinfo {author}
  {\bibfnamefont {I.}~\bibnamefont {Siddiqi}},\ }\bibfield  {title} {\bibinfo
  {title} {Stabilizing entanglement via symmetry-selective bath engineering in
  superconducting qubits},\ }\href
  {https://doi.org/10.1103/PhysRevLett.116.240503} {\bibfield  {journal}
  {\bibinfo  {journal} {Phys. Rev. Lett.}\ }\textbf {\bibinfo {volume} {116}},\
  \bibinfo {pages} {240503} (\bibinfo {year} {2016})}\BibitemShut {NoStop}%
\bibitem [{\citenamefont {Lu}\ \emph {et~al.}(2017)\citenamefont {Lu},
  \citenamefont {Chakram}, \citenamefont {Leung}, \citenamefont {Earnest},
  \citenamefont {Naik}, \citenamefont {Huang}, \citenamefont {Groszkowski},
  \citenamefont {Kapit}, \citenamefont {Koch},\ and\ \citenamefont
  {Schuster}}]{Yao2017}%
  \BibitemOpen
  \bibfield  {author} {\bibinfo {author} {\bibfnamefont {Y.}~\bibnamefont
  {Lu}}, \bibinfo {author} {\bibfnamefont {S.}~\bibnamefont {Chakram}},
  \bibinfo {author} {\bibfnamefont {N.}~\bibnamefont {Leung}}, \bibinfo
  {author} {\bibfnamefont {N.}~\bibnamefont {Earnest}}, \bibinfo {author}
  {\bibfnamefont {R.~K.}\ \bibnamefont {Naik}}, \bibinfo {author}
  {\bibfnamefont {Z.}~\bibnamefont {Huang}}, \bibinfo {author} {\bibfnamefont
  {P.}~\bibnamefont {Groszkowski}}, \bibinfo {author} {\bibfnamefont
  {E.}~\bibnamefont {Kapit}}, \bibinfo {author} {\bibfnamefont
  {J.}~\bibnamefont {Koch}},\ and\ \bibinfo {author} {\bibfnamefont {D.~I.}\
  \bibnamefont {Schuster}},\ }\bibfield  {title} {\bibinfo {title} {Universal
  stabilization of a parametrically coupled qubit},\ }\href
  {https://doi.org/10.1103/PhysRevLett.119.150502} {\bibfield  {journal}
  {\bibinfo  {journal} {Phys. Rev. Lett.}\ }\textbf {\bibinfo {volume} {119}},\
  \bibinfo {pages} {150502} (\bibinfo {year} {2017})}\BibitemShut {NoStop}%
\bibitem [{\citenamefont {Huang}\ \emph {et~al.}(2018)\citenamefont {Huang},
  \citenamefont {Lu}, \citenamefont {Kapit}, \citenamefont {Schuster},\ and\
  \citenamefont {Koch}}]{2018Stabilize}%
  \BibitemOpen
  \bibfield  {author} {\bibinfo {author} {\bibfnamefont {Z.}~\bibnamefont
  {Huang}}, \bibinfo {author} {\bibfnamefont {Y.}~\bibnamefont {Lu}}, \bibinfo
  {author} {\bibfnamefont {E.}~\bibnamefont {Kapit}}, \bibinfo {author}
  {\bibfnamefont {D.~I.}\ \bibnamefont {Schuster}},\ and\ \bibinfo {author}
  {\bibfnamefont {J.}~\bibnamefont {Koch}},\ }\bibfield  {title} {\bibinfo
  {title} {Universal stabilization of single-qubit states using a tunable
  coupler},\ }\href {https://doi.org/10.1103/PhysRevA.97.062345} {\bibfield
  {journal} {\bibinfo  {journal} {Phys. Rev. A}\ }\textbf {\bibinfo {volume}
  {97}},\ \bibinfo {pages} {062345} (\bibinfo {year} {2018})}\BibitemShut
  {NoStop}%
\bibitem [{\citenamefont {Andersen}\ \emph {et~al.}(2019)\citenamefont
  {Andersen}, \citenamefont {Remm}, \citenamefont {Lazar}, \citenamefont
  {Krinner}, \citenamefont {Heinsoo}, \citenamefont {Besse}, \citenamefont
  {Gabureac}, \citenamefont {Wallraff},\ and\ \citenamefont
  {Eichler}}]{Andersen2019}%
  \BibitemOpen
  \bibfield  {author} {\bibinfo {author} {\bibfnamefont {C.~K.}\ \bibnamefont
  {Andersen}}, \bibinfo {author} {\bibfnamefont {A.}~\bibnamefont {Remm}},
  \bibinfo {author} {\bibfnamefont {S.}~\bibnamefont {Lazar}}, \bibinfo
  {author} {\bibfnamefont {S.}~\bibnamefont {Krinner}}, \bibinfo {author}
  {\bibfnamefont {J.}~\bibnamefont {Heinsoo}}, \bibinfo {author} {\bibfnamefont
  {J.-C.}\ \bibnamefont {Besse}}, \bibinfo {author} {\bibfnamefont
  {M.}~\bibnamefont {Gabureac}}, \bibinfo {author} {\bibfnamefont
  {A.}~\bibnamefont {Wallraff}},\ and\ \bibinfo {author} {\bibfnamefont
  {C.}~\bibnamefont {Eichler}},\ }\bibfield  {title} {\bibinfo {title}
  {Entanglement stabilization using ancilla-based parity detection and
  real-time feedback in superconducting circuits},\ }\href
  {https://doi.org/10.1038/s41534-019-0185-4} {\bibfield  {journal} {\bibinfo
  {journal} {npj Quantum Information}\ }\textbf {\bibinfo {volume} {5}},\
  \bibinfo {pages} {69} (\bibinfo {year} {2019})}\BibitemShut {NoStop}%
\bibitem [{\citenamefont {Ma}\ \emph {et~al.}(2019)\citenamefont {Ma},
  \citenamefont {Li}, \citenamefont {Liu}, \citenamefont {Xie},\ and\
  \citenamefont {Li}}]{Fu2019}%
  \BibitemOpen
  \bibfield  {author} {\bibinfo {author} {\bibfnamefont {S.-l.}\ \bibnamefont
  {Ma}}, \bibinfo {author} {\bibfnamefont {X.-k.}\ \bibnamefont {Li}}, \bibinfo
  {author} {\bibfnamefont {X.-y.}\ \bibnamefont {Liu}}, \bibinfo {author}
  {\bibfnamefont {J.-k.}\ \bibnamefont {Xie}},\ and\ \bibinfo {author}
  {\bibfnamefont {F.-l.}\ \bibnamefont {Li}},\ }\bibfield  {title} {\bibinfo
  {title} {Stabilizing bell states of two separated superconducting qubits via
  quantum reservoir engineering},\ }\href
  {https://doi.org/10.1103/PhysRevA.99.042336} {\bibfield  {journal} {\bibinfo
  {journal} {Phys. Rev. A}\ }\textbf {\bibinfo {volume} {99}},\ \bibinfo
  {pages} {042336} (\bibinfo {year} {2019})}\BibitemShut {NoStop}%
\bibitem [{\citenamefont {Bultink}\ \emph {et~al.}(2020)\citenamefont
  {Bultink}, \citenamefont {O’Brien}, \citenamefont {Vollmer}, \citenamefont
  {Muthusubramanian}, \citenamefont {Beekman}, \citenamefont {Rol},
  \citenamefont {Fu}, \citenamefont {Tarasinski}, \citenamefont {Ostroukh},
  \citenamefont {Varbanov}, \citenamefont {Bruno},\ and\ \citenamefont
  {DiCarlo}}]{CC2020}%
  \BibitemOpen
  \bibfield  {author} {\bibinfo {author} {\bibfnamefont {C.~C.}\ \bibnamefont
  {Bultink}}, \bibinfo {author} {\bibfnamefont {T.~E.}\ \bibnamefont
  {O’Brien}}, \bibinfo {author} {\bibfnamefont {R.}~\bibnamefont {Vollmer}},
  \bibinfo {author} {\bibfnamefont {N.}~\bibnamefont {Muthusubramanian}},
  \bibinfo {author} {\bibfnamefont {M.~W.}\ \bibnamefont {Beekman}}, \bibinfo
  {author} {\bibfnamefont {M.~A.}\ \bibnamefont {Rol}}, \bibinfo {author}
  {\bibfnamefont {X.}~\bibnamefont {Fu}}, \bibinfo {author} {\bibfnamefont
  {B.}~\bibnamefont {Tarasinski}}, \bibinfo {author} {\bibfnamefont
  {V.}~\bibnamefont {Ostroukh}}, \bibinfo {author} {\bibfnamefont
  {B.}~\bibnamefont {Varbanov}}, \bibinfo {author} {\bibfnamefont
  {A.}~\bibnamefont {Bruno}},\ and\ \bibinfo {author} {\bibfnamefont
  {L.}~\bibnamefont {DiCarlo}},\ }\bibfield  {title} {\bibinfo {title}
  {Protecting quantum entanglement from leakage and qubit errors via repetitive
  parity measurements},\ }\href {https://doi.org/10.1126/sciadv.aay3050}
  {\bibfield  {journal} {\bibinfo  {journal} {Science Advances}\ }\textbf
  {\bibinfo {volume} {6}},\ \bibinfo {pages} {eaay3050} (\bibinfo {year}
  {2020})},\ \Eprint
  {https://arxiv.org/abs/https://www.science.org/doi/pdf/10.1126/sciadv.aay3050}
  {https://www.science.org/doi/pdf/10.1126/sciadv.aay3050} \BibitemShut
  {NoStop}%
\bibitem [{\citenamefont {Brown}\ \emph {et~al.}(2022)\citenamefont {Brown},
  \citenamefont {Doucet}, \citenamefont {Rist{\`e}}, \citenamefont {Ribeill},
  \citenamefont {Cicak}, \citenamefont {Aumentado}, \citenamefont {Simmonds},
  \citenamefont {Govia}, \citenamefont {Kamal},\ and\ \citenamefont
  {Ranzani}}]{Brown2022}%
  \BibitemOpen
  \bibfield  {author} {\bibinfo {author} {\bibfnamefont {T.}~\bibnamefont
  {Brown}}, \bibinfo {author} {\bibfnamefont {E.}~\bibnamefont {Doucet}},
  \bibinfo {author} {\bibfnamefont {D.}~\bibnamefont {Rist{\`e}}}, \bibinfo
  {author} {\bibfnamefont {G.}~\bibnamefont {Ribeill}}, \bibinfo {author}
  {\bibfnamefont {K.}~\bibnamefont {Cicak}}, \bibinfo {author} {\bibfnamefont
  {J.}~\bibnamefont {Aumentado}}, \bibinfo {author} {\bibfnamefont
  {R.}~\bibnamefont {Simmonds}}, \bibinfo {author} {\bibfnamefont
  {L.}~\bibnamefont {Govia}}, \bibinfo {author} {\bibfnamefont
  {A.}~\bibnamefont {Kamal}},\ and\ \bibinfo {author} {\bibfnamefont
  {L.}~\bibnamefont {Ranzani}},\ }\bibfield  {title} {\bibinfo {title} {Trade
  off-free entanglement stabilization in a superconducting qutrit-qubit
  system},\ }\href {https://doi.org/10.1038/s41467-022-31638-0} {\bibfield
  {journal} {\bibinfo  {journal} {Nature Communications}\ }\textbf {\bibinfo
  {volume} {13}},\ \bibinfo {pages} {3994} (\bibinfo {year}
  {2022})}\BibitemShut {NoStop}%
\bibitem [{\citenamefont {Lin}\ \emph {et~al.}(2013)\citenamefont {Lin},
  \citenamefont {Gaebler}, \citenamefont {Reiter}, \citenamefont {Tan},
  \citenamefont {Bowler}, \citenamefont {S{\o}rensen}, \citenamefont
  {Leibfried},\ and\ \citenamefont {Wineland}}]{Lin2013}%
  \BibitemOpen
  \bibfield  {author} {\bibinfo {author} {\bibfnamefont {Y.}~\bibnamefont
  {Lin}}, \bibinfo {author} {\bibfnamefont {J.~P.}\ \bibnamefont {Gaebler}},
  \bibinfo {author} {\bibfnamefont {F.}~\bibnamefont {Reiter}}, \bibinfo
  {author} {\bibfnamefont {T.~R.}\ \bibnamefont {Tan}}, \bibinfo {author}
  {\bibfnamefont {R.}~\bibnamefont {Bowler}}, \bibinfo {author} {\bibfnamefont
  {A.~S.}\ \bibnamefont {S{\o}rensen}}, \bibinfo {author} {\bibfnamefont
  {D.}~\bibnamefont {Leibfried}},\ and\ \bibinfo {author} {\bibfnamefont
  {D.~J.}\ \bibnamefont {Wineland}},\ }\bibfield  {title} {\bibinfo {title}
  {Dissipative production of a maximally entangled steady state of two quantum
  bits},\ }\href {https://doi.org/10.1038/nature12801} {\bibfield  {journal}
  {\bibinfo  {journal} {Nature}\ }\textbf {\bibinfo {volume} {504}},\ \bibinfo
  {pages} {415} (\bibinfo {year} {2013})}\BibitemShut {NoStop}%
\bibitem [{\citenamefont {Cole}\ \emph {et~al.}(2022)\citenamefont {Cole},
  \citenamefont {Erickson}, \citenamefont {Zarantonello}, \citenamefont {Horn},
  \citenamefont {Hou}, \citenamefont {Wu}, \citenamefont {Slichter},
  \citenamefont {Reiter}, \citenamefont {Koch},\ and\ \citenamefont
  {Leibfried}}]{Leibfried2022}%
  \BibitemOpen
  \bibfield  {author} {\bibinfo {author} {\bibfnamefont {D.~C.}\ \bibnamefont
  {Cole}}, \bibinfo {author} {\bibfnamefont {S.~D.}\ \bibnamefont {Erickson}},
  \bibinfo {author} {\bibfnamefont {G.}~\bibnamefont {Zarantonello}}, \bibinfo
  {author} {\bibfnamefont {K.~P.}\ \bibnamefont {Horn}}, \bibinfo {author}
  {\bibfnamefont {P.-Y.}\ \bibnamefont {Hou}}, \bibinfo {author} {\bibfnamefont
  {J.~J.}\ \bibnamefont {Wu}}, \bibinfo {author} {\bibfnamefont {D.~H.}\
  \bibnamefont {Slichter}}, \bibinfo {author} {\bibfnamefont {F.}~\bibnamefont
  {Reiter}}, \bibinfo {author} {\bibfnamefont {C.~P.}\ \bibnamefont {Koch}},\
  and\ \bibinfo {author} {\bibfnamefont {D.}~\bibnamefont {Leibfried}},\
  }\bibfield  {title} {\bibinfo {title} {Resource-efficient dissipative
  entanglement of two trapped-ion qubits},\ }\href
  {https://doi.org/10.1103/PhysRevLett.128.080502} {\bibfield  {journal}
  {\bibinfo  {journal} {Phys. Rev. Lett.}\ }\textbf {\bibinfo {volume} {128}},\
  \bibinfo {pages} {080502} (\bibinfo {year} {2022})}\BibitemShut {NoStop}%
\bibitem [{\citenamefont {Lloyd}\ \emph {et~al.}(2014)\citenamefont {Lloyd},
  \citenamefont {Mohseni},\ and\ \citenamefont {Rebentrost}}]{Lloyd2014}%
  \BibitemOpen
  \bibfield  {author} {\bibinfo {author} {\bibfnamefont {S.}~\bibnamefont
  {Lloyd}}, \bibinfo {author} {\bibfnamefont {M.}~\bibnamefont {Mohseni}},\
  and\ \bibinfo {author} {\bibfnamefont {P.}~\bibnamefont {Rebentrost}},\
  }\bibfield  {title} {\bibinfo {title} {Quantum principal component
  analysis},\ }\href {https://doi.org/10.1038/nphys3029} {\bibfield  {journal}
  {\bibinfo  {journal} {Nature Physics}\ }\textbf {\bibinfo {volume} {10}},\
  \bibinfo {pages} {631} (\bibinfo {year} {2014})}\BibitemShut {NoStop}%
\bibitem [{\citenamefont {Kjaergaard}\ \emph {et~al.}(2022)\citenamefont
  {Kjaergaard}, \citenamefont {Schwartz}, \citenamefont {Greene}, \citenamefont
  {Samach}, \citenamefont {Bengtsson}, \citenamefont {O'Keeffe}, \citenamefont
  {McNally}, \citenamefont {Braum\"uller}, \citenamefont {Kim}, \citenamefont
  {Krantz}, \citenamefont {Marvian}, \citenamefont {Melville}, \citenamefont
  {Niedzielski}, \citenamefont {Sung}, \citenamefont {Winik}, \citenamefont
  {Yoder}, \citenamefont {Rosenberg}, \citenamefont {Obenland}, \citenamefont
  {Lloyd}, \citenamefont {Orlando}, \citenamefont {Marvian}, \citenamefont
  {Gustavsson},\ and\ \citenamefont {Oliver}}]{2022matrix}%
  \BibitemOpen
  \bibfield  {author} {\bibinfo {author} {\bibfnamefont {M.}~\bibnamefont
  {Kjaergaard}}, \bibinfo {author} {\bibfnamefont {M.~E.}\ \bibnamefont
  {Schwartz}}, \bibinfo {author} {\bibfnamefont {A.}~\bibnamefont {Greene}},
  \bibinfo {author} {\bibfnamefont {G.~O.}\ \bibnamefont {Samach}}, \bibinfo
  {author} {\bibfnamefont {A.}~\bibnamefont {Bengtsson}}, \bibinfo {author}
  {\bibfnamefont {M.}~\bibnamefont {O'Keeffe}}, \bibinfo {author}
  {\bibfnamefont {C.~M.}\ \bibnamefont {McNally}}, \bibinfo {author}
  {\bibfnamefont {J.}~\bibnamefont {Braum\"uller}}, \bibinfo {author}
  {\bibfnamefont {D.~K.}\ \bibnamefont {Kim}}, \bibinfo {author} {\bibfnamefont
  {P.}~\bibnamefont {Krantz}}, \bibinfo {author} {\bibfnamefont
  {M.}~\bibnamefont {Marvian}}, \bibinfo {author} {\bibfnamefont
  {A.}~\bibnamefont {Melville}}, \bibinfo {author} {\bibfnamefont {B.~M.}\
  \bibnamefont {Niedzielski}}, \bibinfo {author} {\bibfnamefont
  {Y.}~\bibnamefont {Sung}}, \bibinfo {author} {\bibfnamefont {R.}~\bibnamefont
  {Winik}}, \bibinfo {author} {\bibfnamefont {J.}~\bibnamefont {Yoder}},
  \bibinfo {author} {\bibfnamefont {D.}~\bibnamefont {Rosenberg}}, \bibinfo
  {author} {\bibfnamefont {K.}~\bibnamefont {Obenland}}, \bibinfo {author}
  {\bibfnamefont {S.}~\bibnamefont {Lloyd}}, \bibinfo {author} {\bibfnamefont
  {T.~P.}\ \bibnamefont {Orlando}}, \bibinfo {author} {\bibfnamefont
  {I.}~\bibnamefont {Marvian}}, \bibinfo {author} {\bibfnamefont
  {S.}~\bibnamefont {Gustavsson}},\ and\ \bibinfo {author} {\bibfnamefont
  {W.~D.}\ \bibnamefont {Oliver}},\ }\bibfield  {title} {\bibinfo {title}
  {Demonstration of density matrix exponentiation using a superconducting
  quantum processor},\ }\href {https://doi.org/10.1103/PhysRevX.12.011005}
  {\bibfield  {journal} {\bibinfo  {journal} {Phys. Rev. X}\ }\textbf {\bibinfo
  {volume} {12}},\ \bibinfo {pages} {011005} (\bibinfo {year}
  {2022})}\BibitemShut {NoStop}%
\bibitem [{\citenamefont {Lu}(2019)}]{yaothesis2019}%
  \BibitemOpen
  \bibfield  {author} {\bibinfo {author} {\bibfnamefont {Y.}~\bibnamefont
  {Lu}},\ }\href {https://doi.org/10.6082/uchicago.1973} {\bibinfo {title}
  {Parametric control of flux-tunable superconducting circuits}} (\bibinfo
  {year} {2019})\BibitemShut {NoStop}%
\bibitem [{\citenamefont {Roy}\ \emph {et~al.}(2023)\citenamefont {Roy},
  \citenamefont {Li}, \citenamefont {Kapit},\ and\ \citenamefont
  {Schuster}}]{roy2022realization}%
  \BibitemOpen
  \bibfield  {author} {\bibinfo {author} {\bibfnamefont {T.}~\bibnamefont
  {Roy}}, \bibinfo {author} {\bibfnamefont {Z.}~\bibnamefont {Li}}, \bibinfo
  {author} {\bibfnamefont {E.}~\bibnamefont {Kapit}},\ and\ \bibinfo {author}
  {\bibfnamefont {D.}~\bibnamefont {Schuster}},\ }\bibfield  {title} {\bibinfo
  {title} {Two-qutrit quantum algorithms on a programmable superconducting
  processor},\ }\href {https://doi.org/10.1103/PhysRevApplied.19.064024}
  {\bibfield  {journal} {\bibinfo  {journal} {Phys. Rev. Appl.}\ }\textbf
  {\bibinfo {volume} {19}},\ \bibinfo {pages} {064024} (\bibinfo {year}
  {2023})}\BibitemShut {NoStop}%
\bibitem [{\citenamefont {Kapit}\ \emph {et~al.}(2014)\citenamefont {Kapit},
  \citenamefont {Hafezi},\ and\ \citenamefont {Simon}}]{Eliot2014}%
  \BibitemOpen
  \bibfield  {author} {\bibinfo {author} {\bibfnamefont {E.}~\bibnamefont
  {Kapit}}, \bibinfo {author} {\bibfnamefont {M.}~\bibnamefont {Hafezi}},\ and\
  \bibinfo {author} {\bibfnamefont {S.~H.}\ \bibnamefont {Simon}},\ }\bibfield
  {title} {\bibinfo {title} {Induced self-stabilization in fractional quantum
  hall states of light},\ }\href {https://doi.org/10.1103/PhysRevX.4.031039}
  {\bibfield  {journal} {\bibinfo  {journal} {Phys. Rev. X}\ }\textbf {\bibinfo
  {volume} {4}},\ \bibinfo {pages} {031039} (\bibinfo {year}
  {2014})}\BibitemShut {NoStop}%
\bibitem [{\citenamefont {Roy}\ \emph {et~al.}(2021)\citenamefont {Roy},
  \citenamefont {Li}, \citenamefont {Kapit},\ and\ \citenamefont
  {Schuster}}]{ZZcorrection}%
  \BibitemOpen
  \bibfield  {author} {\bibinfo {author} {\bibfnamefont {T.}~\bibnamefont
  {Roy}}, \bibinfo {author} {\bibfnamefont {Z.}~\bibnamefont {Li}}, \bibinfo
  {author} {\bibfnamefont {E.}~\bibnamefont {Kapit}},\ and\ \bibinfo {author}
  {\bibfnamefont {D.~I.}\ \bibnamefont {Schuster}},\ }\href@noop {} {\bibinfo
  {title} {Tomography in the presence of stray inter-qubit coupling}} (\bibinfo
  {year} {2021}),\ \Eprint {https://arxiv.org/abs/2103.13611}
  {arXiv:2103.13611} \BibitemShut {NoStop}%
\bibitem [{\citenamefont {li~Ma}\ \emph {et~al.}(2021)\citenamefont {li~Ma},
  \citenamefont {Zhang}, \citenamefont {ke~Li}, \citenamefont {long Ren},
  \citenamefont {kun Xie}, \citenamefont {tao Cao},\ and\ \citenamefont
  {li~Li}}]{Ma_2021}%
  \BibitemOpen
  \bibfield  {author} {\bibinfo {author} {\bibfnamefont {S.}~\bibnamefont
  {li~Ma}}, \bibinfo {author} {\bibfnamefont {J.}~\bibnamefont {Zhang}},
  \bibinfo {author} {\bibfnamefont {X.}~\bibnamefont {ke~Li}}, \bibinfo
  {author} {\bibfnamefont {Y.}~\bibnamefont {long Ren}}, \bibinfo {author}
  {\bibfnamefont {J.}~\bibnamefont {kun Xie}}, \bibinfo {author} {\bibfnamefont
  {M.}~\bibnamefont {tao Cao}},\ and\ \bibinfo {author} {\bibfnamefont
  {F.}~\bibnamefont {li~Li}},\ }\bibfield  {title} {\bibinfo {title}
  {Coupling-modulation–mediated generation of stable entanglement of
  superconducting qubits via dissipation},\ }\href
  {https://doi.org/10.1209/0295-5075/ac2b5c} {\bibfield  {journal} {\bibinfo
  {journal} {Europhysics Letters}\ }\textbf {\bibinfo {volume} {135}},\
  \bibinfo {pages} {63001} (\bibinfo {year} {2021})}\BibitemShut {NoStop}%
\bibitem [{\citenamefont {Wallraff}\ \emph {et~al.}(2007)\citenamefont
  {Wallraff}, \citenamefont {Schuster}, \citenamefont {Blais}, \citenamefont
  {Gambetta}, \citenamefont {Schreier}, \citenamefont {Frunzio}, \citenamefont
  {Devoret}, \citenamefont {Girvin},\ and\ \citenamefont
  {Schoelkopf}}]{wallraff2007sidband}%
  \BibitemOpen
  \bibfield  {author} {\bibinfo {author} {\bibfnamefont {A.}~\bibnamefont
  {Wallraff}}, \bibinfo {author} {\bibfnamefont {D.~I.}\ \bibnamefont
  {Schuster}}, \bibinfo {author} {\bibfnamefont {A.}~\bibnamefont {Blais}},
  \bibinfo {author} {\bibfnamefont {J.~M.}\ \bibnamefont {Gambetta}}, \bibinfo
  {author} {\bibfnamefont {J.}~\bibnamefont {Schreier}}, \bibinfo {author}
  {\bibfnamefont {L.}~\bibnamefont {Frunzio}}, \bibinfo {author} {\bibfnamefont
  {M.~H.}\ \bibnamefont {Devoret}}, \bibinfo {author} {\bibfnamefont {S.~M.}\
  \bibnamefont {Girvin}},\ and\ \bibinfo {author} {\bibfnamefont {R.~J.}\
  \bibnamefont {Schoelkopf}},\ }\bibfield  {title} {\bibinfo {title} {Sideband
  transitions and two-tone spectroscopy of a superconducting qubit strongly
  coupled to an on-chip cavity},\ }\href
  {https://doi.org/10.1103/PhysRevLett.99.050501} {\bibfield  {journal}
  {\bibinfo  {journal} {Phys. Rev. Lett.}\ }\textbf {\bibinfo {volume} {99}},\
  \bibinfo {pages} {050501} (\bibinfo {year} {2007})}\BibitemShut {NoStop}%
\end{thebibliography}%
\end{document}